\documentclass[useAMS,usenatbib,onecolumn]{mn2e}
\usepackage{graphicx}
\usepackage{amsmath}
\usepackage{array}
\usepackage{multicol}
\usepackage{multirow}
\usepackage{longtable}
\usepackage{url}
\usepackage[colorlinks=true,
            linkcolor=blue,
            urlcolor=blue,
            citecolor=blue]{hyperref}
\usepackage{pdflscape}

\title[Dilution factors of SE-SNe: Derivation]{Empirically determined dilution factors of stripped-envelope, core-collapse SNe: Paper I - Method \& Progenitor constraints}

\author[Cano]{\noindent Zach Cano$^{1,2,3}$\thanks{zewcano@gmail.com}  \\
\noindent $^1$Centre for Astrophysics and Cosmology, Science Institute, University of Iceland, Dunhagi 5, 107 Reykjavik, Iceland.\\
\noindent $^2$Instituto de Astrof\'isica de Andaluc\'ia (IAA-CSIC), Glorieta de la Astronom\'ia s/n, E-18008, Granada, Spain.\\
\noindent $^3$Juan de la Cierva Fellow.\\
}

\begin{document}

\date{Accepted xx. Received xx; in original form xx}

\pagerange{\pageref{firstpage}--\pageref{lastpage}} \pubyear{2013}

\maketitle

\label{firstpage}

\begin{abstract}
In this work, the empirically derived dilution/correct factors of a sample stripped-envelope, core-collapse supernovae (SE-SNe), including five SNe IIb, four SNe Ib, six SNe Ic and two relativistic broad-lined type Ic supernovae (SNe IcBL) are presented.  The ultimate goal of this project is to derive model-free distances to the host galaxy of one or more gamma-ray burst supernova (GRB-SN), and to exploit their observed luminosity$-$decline relationship by employing them as cosmological probes.  In the first part of a two-paper analysis, I present my method for deriving the dilution factors of the SE-SN sample, which were chosen on the basis that cosmological-model-independent distances exist to their host galaxies, and each has a sufficient dataset that allows for host-subtracted, dereddened rest-frame $BVI$ LCs to be constructed, and time-series spectra.  A Planck function was fit to the data to derive the blackbody radius and blackbody temperature as a function of time, while the blueshifted velocity of either Si \textsc{ii}~$\lambda$6355 or Fe~\textsc{ii}~$\lambda$5169 was used a proxy of the photospheric velocity, and hence photospheric radius.  The ratio of these empirically derived radii was taken as the dilution/correct factor. I then compared the empirically derived dilution factors with synthetic values obtained from radiative transfer models calculated for SE-SNe arising from binary systems.  It is seen that the empirical dilution factors of the SNe Ic and GRB-SNe, the latter which were derived based on luminosity distances calculated from their spectroscopic redshift, are very similar. It is found that the dilution factors of the two relativistic SN IcBL are very different to those of the GRB-SNe, meaning that these engine-driven events may arise from fundamentally different progenitor systems.
\end{abstract}


\section{Introduction}

Stripped-envelope supernovae (SE-SNe) arise from the core-collapse of massive stars.  Based on phenomenology classifications \citep{Filippenko97}, all type I SNe are devoid of hydrogen features in their optical spectra, where SNe Ib display helium features that are absent in the spectra of SNe Ic.  SNe IIb show weak hydrogen features in their optical spectra before maximum, which disappear and replaced by helium absorption lines.  The transition from SNe IIb $\rightarrow$ Ib $\rightarrow$ Ic implies greater degrees of envelope stripping experienced by their progenitors stars prior to explosion: for example, it is thought that SNe Ic their outer hydrogen and helium envelopes have been stripped away completely, hence the absence of these lines in their observed spectra.

The progenitors of SE-SNe are massive stars whose initial zero-age main sequence (ZAMS) masses are larger than those attributed to the progenitors of SNe II.  Circumstantial evidence for this supposition arises from statistical analyses of the environments of SE-SNe, where it has been seen that, on average, SNe Ib and Ic occur in the brightest (and thus most star-forming) regions of their host galaxies \citep{Fruchter06,Kelly08} relative to SNe II \citep{AndersonJames08}.  As massive stars are thought to have cosmologically short lifespans (of order a few to a few tens of millions of years), they are not expected to travel very far from their formation region. This is in contrast to less massive stars that live longer and  may travel much further from their regions of origin, and thus display less association with star-forming regions. While several progenitors of SNe II have been identified in archival images (e.g. \citealt{Smartt09}), including SNe IIb (e.g. \citealt{Aldering94,Maund11,Kilpatrick17}) initial searches for SNe Ib/Ic resulted only in deep upper-limits \citep{Eldridge13}.  A candidate stellar system, consisting either a single Wolf-Rayet with a ZAMS mass of $30\sim35$~M$_{\odot}$ \citep{Groh13} or a less-massive ($3.5-11.0$~M$_{\odot}$) Wolf-Rayet star in a binary system \citep{Bersten14} was proposed for type Ib SN iPTF13bvn \citep{Cao13}, although the association is tentative as the candidate was located $\approx$2$\sigma$ from the position of the SN.  Recently, a bright, blue star was found in Hubble Space Telescope imaging at the position of type Ic SN~2017ein by \citet{VanDyk17}, suggesting the first direct observation of the progenitor star of a highly stripped core-collapse SN.

Additional, indirect constraints on the progenitor stars of SE-SNe comes from modelling photometric and spectroscopic observations of the SNe themselves: sophisticated hydrodynamical models can be used to infer the mass, radius and chemistry of the exploded star, while radiative transfer models can help constrain the physical and chemical properties of the SN's atmosphere.  In turn, analytical models can also give useful and complementary constraints on the SNe themselves, providing a first-order approximation of their bolometric (ejecta mass and nickel content therein, and kinetic energy) properties.   Statistical analyses based on the results of fitting analytical models to bolometric light-curves (LCs) of SE-SNe have found that, on average, SNe Ib and Ib have similar ejecta and nickel masses and kinetic energies, where broad-lined type Ic (SNe IcBL), including the SNe associated with gamma-ray bursts, i.e. GRB-SNe \citep{WoosleyBloom06,CanoReview17}, have larger ejecta and nickel masses and kinetic energies \citep{Cano13,Taddia15,Lyman16,Prentice16}.  Conversely, SNe IIb have smaller kinetic energies and ejecta/nickel masses \citep{Lyman16}.  Under the assumption that all SE-SNe arise from single, massive stars, these results suggest SNe Ib and Ic arise from stars with similar ZAMS masses, while those of SNe IIb are less massive, and those of GRB-SNe are more massive. Indeed, when one considers that increased progenitor mass implies more mass-loss via line-driven stellar winds or nuclear burning instabilities prior to explosion, these results appear logical.  However, the role of binarity cannot be ignored, which will result in, among other things, reduced progenitor masses as binary interactions can also efficiently strip stellar envelopes before explosion.

The observational properties of SE-SNe are also of great interest to the research community. In 2014 it was demonstrated that GRB-SNe have observed relationships between their absolute peak brightness and the shape of their optical LCs: a luminosity$-$stretch relation \citep{Cano14} and an analogous luminosity$-$decline relation (LDR; \citealt{CJG14}, CJG14 hereafter; \citealt{LiHjorth14}).  It was seen that two relativistic SNe IcBL, SN~2009bb \citep{Soderberg10,Pignata2011} and SN~2012ap \citep{Margutti14,Chakraborti15}, which like GRB-SNe are thought to be engine-driven SNe, follow the same LDR as GRB-SNe (CJG14), which provided additional indirect arguments that the progenitors of GRB-SNe and relativistic SNe IcBL may share some physical similarities.  The amount of scatter in the GRB-SNe/relativistic SNe IcBL $BVR$ LDRs was of order $\sigma = 0.2-0.3$~magnitudes.  Conversely, it was shown that SNe IIb, Ib, Ic and type Ic superluminous supernovae (SLSNe-Ic) do not have a LDR (CJG14)\footnote{Though see \citet{InserraSmartt14} for a counter-argument regarding SLSNe-Ic.}.

A major hurdle that needs to be overcome to successfully facilitate the use of GRB-SNe as cosmological probes is to determine distances to their host galaxies in a manner that is entirely independent of any cosmological model.  There are many ways to achieve this goal, with the most relevant choices being (1) detecting and monitoring Cepheid variable stars in the host galaxy of the nearest GRB-SNe, SN~1998bw, (2) using the Tully-Fisher relationship of the same galaxy, or (3) using a kinematic model, e.g. the Expanding Photosphere Method (EPM).  Using current technology, option (1) is observationally expensive with only a modest scientific return, while it was shown that the host galaxy of SN~1998bw does not follow the TF relationship \citep{Arabsalmani15}.  Instead, the EPM, which is a variant of the Baade$-$Wesselink method (Baade 1927), holds a lot of promise.  In this framework, one compares the angular size of the photosphere of a SN with its measured expansion velocity, both of which are determined empirically under the assumption that a SN is a blackbody emitter and its spectral energy distribution (SED) can be modelled and fit with a blackbody/Planck function.  One major constituent of the EPM is knowledge of the dilution factor, which considers that the radius in which the blackbody photons are emitted (the thermalization radius) is not at the same spatial location as the photosphere.  Thus, to effectively use the EPM to determine the distance to a given SN, knowledge of its dilution factor, which depends on the SN's chemistry and ionization state, is required.  

It is the goal of this paper, which is the first of a two-part series, to empirically derive the dilution factors of SE-SNe and use them as a proxy for GRB-SNe to ultimately derive model-free distances to one or more GRB-SN host galaxy and facilitate their use as cosmological probes to determine the Hubble constant in the local Universe. GRB-SN cosmology will form the premise of the analysis in Paper II.  In this current work, Paper I, I will present the method for determining the dilution factors (Section \ref{sec:empirical_dilution_factor}) of SE-SNe whose distances are known independent of any cosmological model.  Model-free distances are absolutely vital for this analysis, so that when once uses the derived dilution factors, they do not introduce any biases into subsequent cosmological studies based upon them.  Moreover, the legacy sample of dilution factors presented here can also provide constraints on their progenitors themselves, and be used to calibrate radiative transfer models accordingly. Hence, in Section \ref{sec:discussionII_RT_compare} I compare the derived dilution factors with synthetic dilution factors derived from radiative transfer models of SE-SNe arising from binary systems (\citealt{Dessart15}; D15 hereafter).  This discussion follows an inter-comparison of the different SE-SN subtypes in Section~\ref{sec:discussionI_intercompare}.  In Section~\ref{sec:discussionIII_GRBSNe} I present empirical dilution factors of GRB-SNe that are calculated using luminosity distances calculated based upon a generic cosmology, and compare these with the model-free dilution factors of the SE-SNe and those from the radiative transfer model, drawing general conclusions of the physical properties of GRB-SNe relative to SE-SNe.  In Section~\ref{sec:Caveats} I discuss the limitations of my approach, and finally in Section~\ref{sec:conclusions} I summarise the work.

\begin{table*}
\small
\centering
\setlength{\tabcolsep}{5pt}
\caption{SE-SNe: Vital statistics}
\label{table:vitals}
\begin{tabular}{|ccccccc|}
\hline
SN	&	Type	&	$z$	&	$t_0$ (JD)	&	$E(B-V)_{\rm fore}$ (mag)	&	$E(B-V)_{\rm host}$ (mag)	&	Ref(s).	\\
\hline													
1994I	&	Ic	&	0.00155	&	$2449440.25\pm1.25$	&	0.03	&	0.42	&	(1$-$19)	\\
2002ap	&	Ic	&	0.002187	&	2452303.4	&	0.06	&	0.03	&	(19, 22$-$37)	\\
2004aw	&	Ic	&	0.0175	&	$2453081.45\pm2.95$	&	0.02	&	0.35	&	(42$-$44)	\\
2005ek	&	Ic	&	0.016618	&	$2453634.6\pm0.4$	&	0.18	&	0	&	(44$-$45)	\\
2007gr	&	Ic	&	0.001729	&	$2454325.5\pm2.5$	&	0.05	&	0.03	&	(49$-$57)	\\
2011bm	&	Ic	&	0.0221	&	$2455645.0\pm1.5$	&	0.03	&	0.03	&	(67)	\\
2009bb	&	IcBL	&	0.009987	&	$2454909.6\pm0.6$	&	0.08	&	$0.50\pm0.07$	&	(44,58)	\\
2012ap	&	IcBL	&	0.012241	&	$2455964.1\pm1.1$	&	0.04	&	$0.83\pm.12$	&	(68$-$69)	\\
1998bw	&	GRB	&	0.00867	&	2450929.409	&	0.05	&	variable	&	(20$-$21)	\\
2003dh	&	GRB	&	0.1685	&	2452727.984	&	0.03	&	0.12	&	(38$-$39)	\\
2003lw	&	GRB	&	0.10536	&	2452977.418	&	0.9	&	$0.24\pm.10$	&	(40$-$41)	\\
2006aj	&	GRB	&	0.03342	&	2453787.649	&	0.03	&	$0.05\pm0.01$	&	(46$-$48)	\\
2009nz	&	GRB	&	0.49	&	2455163.476	&	0.03	&	0	&	(59$-$63)	\\
2010bh	&	GRB	&	0.0592	&	2455272.031	&	0.1	&	$0.16\pm0.01$	&	(64$-$66)	\\
2012bz	&	GRB	&	0.283	&	2456039.800	&	0.03	&	0	&	(70)	\\
2013dx	&	GRB	&	0.145	&	2456475.504	&	0.04	&	0	&	(71$-$72)	\\
2016jca	&	GRB	&	0.1475	&	2457742.280	&	0.03	&	$0.02\pm0.01$	&	(73)	\\
2008ax	&	IIB	&	0.0021	&	$2454528.80\pm0.15$	&	0.02	&	$0.48\pm0.10$	&	(74$-$75)	\\
2010as	&	IIB	&	0.007354	&	$2455271.25\pm3.45$	&	0.15	&	$0.42\pm0.10$	&	(76)	\\
2011dh	&	IIB	&	0.00155	&	2455713	&	0.03	&	0	&	(77)	\\
2011ei	&	IIB	&	0.009317	&	2455767.5	&	0.05	&	0.18	&	(78)	\\
2011hs	&	IIB	&	0.005701	&	$2455872.0\pm4.0$	&	0.01	&	$0.16\pm0.08$	&	(79)	\\
1999dn	&	Ib	&	0.00938	&	$2451408.0\pm2.0$	&	0.05	&	0.05	&	(80$-$81)	\\
2005bf	&	Ib	&	0.018913	&	2453459.5	&	0.04	&	0	&	(82$-$83)	\\
2008D	&	Ib	&	0.0070	&	2454475.06	&	0.02	&	$0.60\pm0.10$	&	(84)	\\
2009jf	&	Ib	&	0.007942	&	$2455099.5\pm1.0$	&	0.1	&	0	&	(85)	\\
\hline
\end{tabular}
\begin{flushleft}
NB: (1) Explosion times are in Julian dates. (2) Extinctions are in units of magnitudes.\\
\scriptsize{\textbf{References}: (1) \cite{Tully1988}; (2) \cite{Baron1996}; (3) \cite{Baron2007}; (4) \cite{Bose2014}; (5) \cite{Chiba1995}; (6) \cite{Ciardullo2002} ; (7) \cite{Dessart2008}; (8) \cite{Feldmeier1997} ; (9) \cite{Ferrarese2000}; (10) \cite{Iwamoto1994} ; (11) \cite{Poznanski2009}; (12) \cite{Richmond1996}; 
(13) \cite{Sauer2009}; (14) \cite{Sofue1994} ; (15) \cite{Takats2006}; (16) \cite{Tonry2001} ; (17) \cite{Tutui1997}; (18) \cite{Vinko2012}; (19) \cite{Zasov1996} ; (20) \cite{Patat01}; (21) \cite{Cloc2011}; (22) \cite{Foley2003}; (23) \cite{GalYam2002}; (24) \cite{Hendry2005}; (25) \cite{Herrmann2008}; (26) \cite{Jang2014} ; (27) \cite{Kinugasa2002}; (28) \cite{Mazzali2002} ; (29) \cite{Olivares2010} ; (30) \cite{Pandey2003}; (31) \cite{Sharina1996}; (32) \cite{Sohn1996}; (33) \cite{Tomita2006}; (35) \cite{VanDyk2006}; (36) \cite{Vinko2004} ; (37) \cite{Yoshii2003} ; (38) \cite{Deng05} ; (39) \cite{Hjorth2003}; (40) \cite{Malesani04}; (41) \cite{Mazzali2006}; (42) \cite{Boles2004}; (43) \cite{Taubenberger2006}; (44) \cite{Theureau2007} ; (45) \cite{Drout2013}; (46) \cite{Sollerman06}; (47) \cite{Ferrero06}; (48) \cite{Pian06}; (49) \cite{Hunter2009}; (50) \cite{Kirshner1974}; (51) \cite{Pierce1994} ; (52) \cite{Schmidt1992}; (53) \cite{Schmidt1994}; (54) \cite{Springob2009} ; (56) \cite{Valenti2008}; (57) \cite{Zinn2011}; (58) \cite{Pignata2011} ; (59) \cite{Berger2011}; (60) \cite{Filgas2011} ; (61) \cite{Cobb2010}; (62) \cite{Vergani2011} ; (63) \cite{Troja2012}; (64) \cite{Olivares2012}; (65) \cite{Bufano2012}; (66) \cite{Cano11b} ; (67) \cite{Valenti2012}; (68) \cite{Liu2015} ; (69) \cite{Milisavljevic2015}; (70) \cite{Schulze2014} ; (71) \cite{Delia2015}; (72) \cite{Toy2016}; (73) \cite{Cano17}; (74) \cite{Taubenberger11}; (75) \cite{Pastorello08}}; (76) \cite{Folatelli14}; (77) \cite{Ergon14}; (78) \cite{Milisavljevic13}; (79) \cite{Bufano14}; (80) \cite{Benetti11}; (81) \cite{CMS14}; (82) \cite{Monard05}; (83) \cite{Anupama05}; (84) \cite{Malesani09}; (85) \cite{Valenti11} \\
\end{flushleft}
\end{table*}

\begin{figure*}
 \centering
 \includegraphics[width=\hsize]{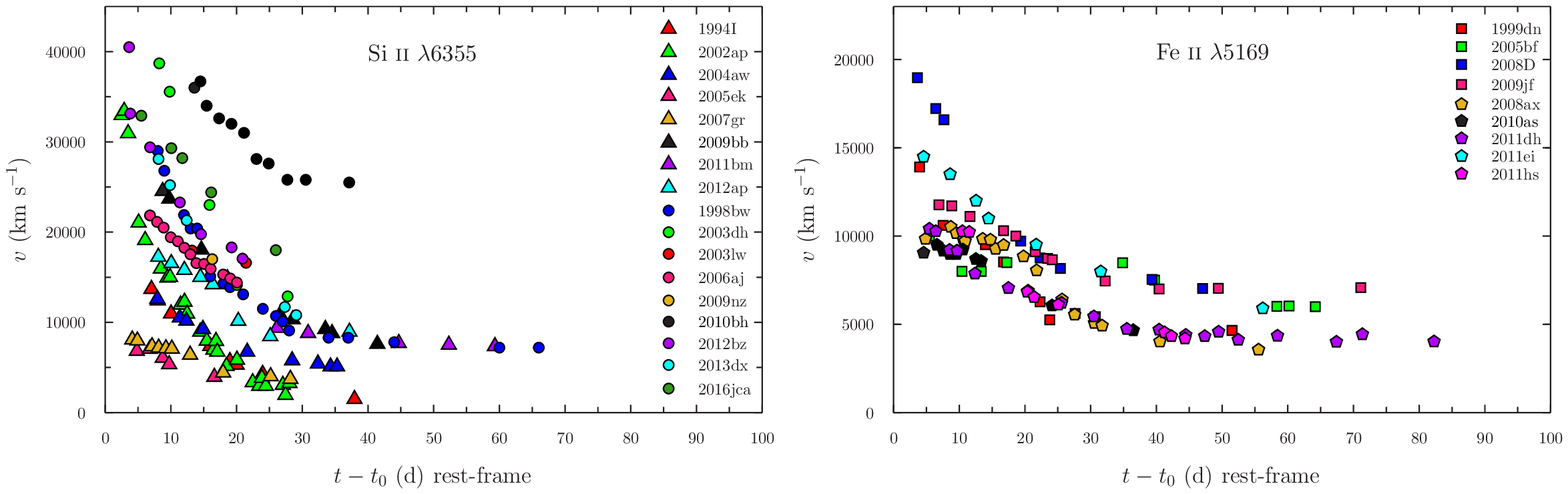} 
 \caption{Line velocities of the sample of SE-SNe. Time are given in the rest-frame relative to the explosion epoch.  Each line velocity is used as a proxy for the photospheric velocity ($v_{\rm ph}$), whereby the photospheric radius is calculated as $R_{\rm phot} = v_{\rm ph}t$. \textit{Left}: Si \textsc{ii} $\lambda$6355 line transition velocities of the SNe Ic in the sample.  It can be seen that GRB-SNe (circles) are more tightly clustered than the SNe Ic (triangles), with the exception of SN~2010bh, which is clearly more rapid than all the other SNe Ic in the sample at a given moment in time. \textit{Right}: Fe \textsc{ii} $\lambda$5169 line transition velocities of the SNe IIb (pentagons) and Ib (squares) in the sample.  Literature references for the velocities of each SE-SN is found in Table \ref{table:vitals}.}
 \label{fig:vels}
\end{figure*}

\begin{figure*}
 \centering
 \includegraphics[width=\hsize]{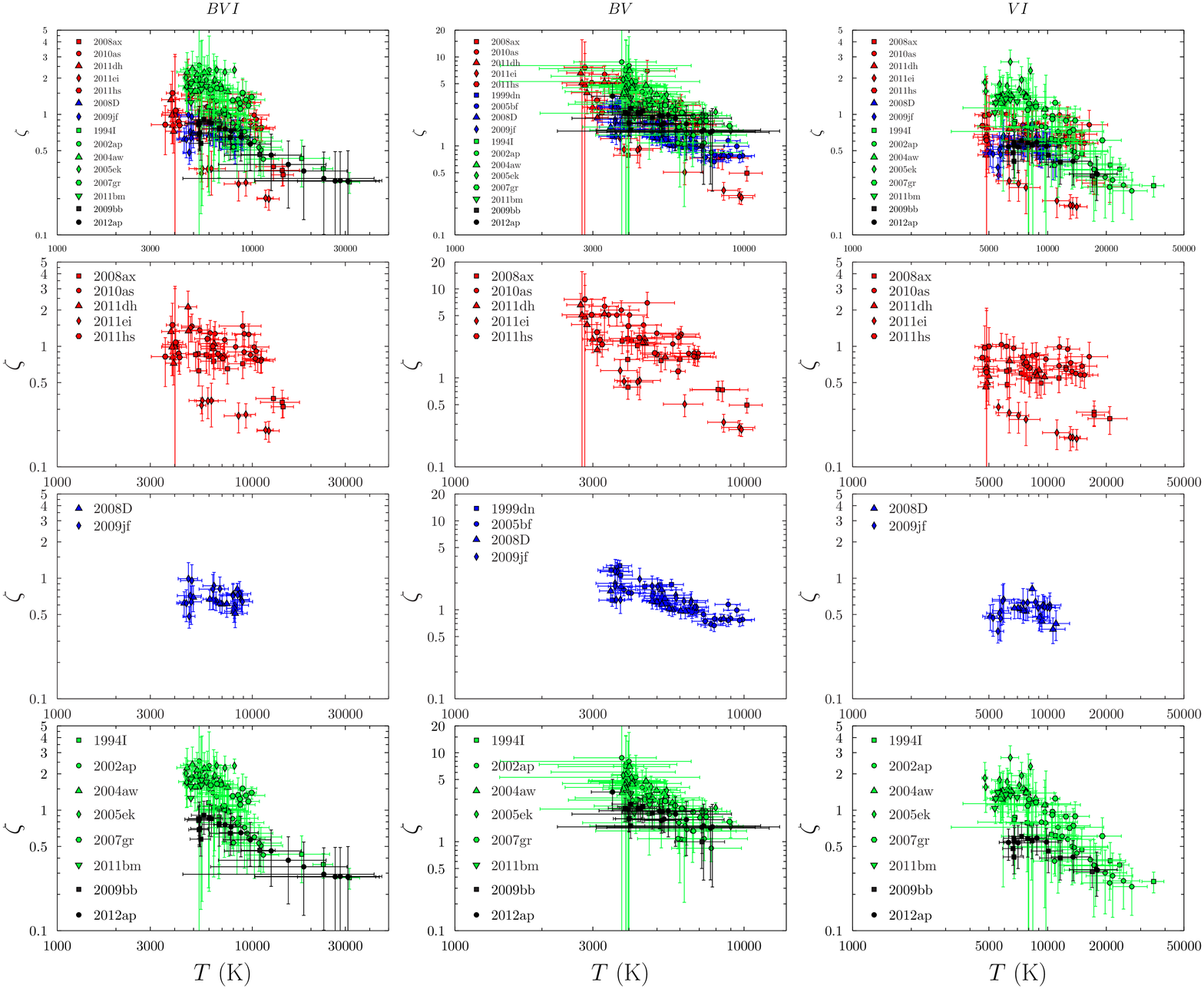} 
 \caption{\textit{Top row}: Empirically derived dilution factors ($\zeta$) in filters ($BVI$, $BV$ and $VI$, left to right) of SNe IIb (red), SNe Ib (blue), SNe Ic (green) and the two relativistic SNe IcBL (black) in the sample as a function of temperature. Each sub-type is respectively presented in the following rows. It is seen that the largest amount of scatter in seen in the $BV$ filter combination, where the dilution factors of the entire SE-SN sample span roughly an order of magnitude at a given temperature.}
 \label{fig:DF_all}
\end{figure*}

\section{Empirically derived dilution factors}
\label{sec:empirical_dilution_factor}

SNe are not perfect blackbody (BB) emitters.  In reality, the thermalization radius is not the same as the photospheric radius (defined as the location where the total inward-integrated radial optical depth reaches a value of $2/3$, e.g. \citealt{DH05}) where the photons escape into space unimpeded.  This is especially so during the photospheric phase when the SN ejecta is partially or near fully ionized and electron-scattering is a significant contributor to the optical opacity, the radius of thermalized outflow layer is less than the photospheric layer ($R_{\rm BB} < R_{\rm phot}$).  This implies that there is a global source of thermalized photon dilution, which historically has been called the dilution factor ($\zeta$) or the distance-correction factor. The physics underlying $\zeta$ is complex, and its precise value depends on the temperature, composition and density of the SN atmosphere, as well as the thermalization radius.  Indeed a key quantity that regulates the amount of dilution present is the spatial separation between $R_{\rm BB}$ and $R_{\rm phot}$, which is the focus of this work here.  Empirically, we can find $\zeta$ for a given SN if we know its explosion time ($t_0$) and its distance from Earth.  Then, $\zeta$ is taken the ratio of $R_{\rm BB}$ to $R_{\rm phot}$:

\begin{equation}
 \zeta = \frac{R_{\rm BB}}{R_{\rm phot}}
\end{equation}

Strictly speaking, $\zeta$ represents the amount of correction needed to transform the measured blackbody flux into the observed flux.  So while the varying photospheric and blackbody radii contribute to this correction term, other factors are also at play including the dilution of flux arising from the strongly scattering SN atmosphere.  So what is referred to here as the dilution factor can be more appropriately referred to as a blackbody$-$observed flux ``correction'' factor.  Nevertheless, the former term is adopted throughout this paper.

The SE-SNe considered here were chosen as cosmological-model-independent distances are known to their host galaxies (Table \ref{table:distances}), they have broadband observations which are host-subtracted, there is knowledge of the entire line-of-sight extinction, they have time-series spectra, and there is an estimate of their explosion date ($t_0$; Table \ref{table:vitals}).  To determine the distance to each SE-SN, I used the distances tabulated in the NASA/IPAC Extragalactic Database (NED\footnote{\url{http://ned.ipac.caltech.edu/}}), and taken the weighted average. Ideally, we would like to use a single line transition as the proxy for the photospheric velocity, where historically both Si~\textsc{ii} $\lambda$6355 of Fe~\textsc{ii} $\lambda$5169 have been used.  One should keep in mind the caveat of using a single transition as the proxy for the photospheric velocity, as the photosphere is unlikely to be a sharp boundary in space, but instead has a certain (unknown, and varying from SN to SN) spatial extent.  However, it is hoped that by using a single transition as a proxy of the photospheric velocity, any unknown systematic uncertainties will affect all SNe in the same manner.  Unfortunately, it is not possible to use a single transition for all of the SE-SNe in the sample: SNe IIb and Ib do not usually have Si~\textsc{ii}~$\lambda$6355 features, while Fe~\textsc{ii} $\lambda$5169 is not always possible to detect in the GRB-SNe in the sample.  Hence, as the ultimate goal is to calibrate the GRB-SNe against the SNe Ic in the sample, I have used the Si~\textsc{ii}~$\lambda$6355 for the SNe Ic, including the IcBL and GRB-SNe.  Instead, Fe~\textsc{ii} $\lambda$5169 is used for the SNe IIb and Ib in the sample (except for SN~2008D, for which I used Fe~\textsc{ii}~$\lambda$5197 from \citealt{Malesani09}), which may introduce some systematic differences when comparing SNe IIb/Ib with SNe Ic/IcBL/GRB-SNe.

The general procedure undertaken to derive the dilution factors of the sample of SE-SNe is:

\begin{enumerate}
 \item Obtain de-reddened (both foreground and local to the SN) and host-subtracted optical observations of each SN in Johnson/Cousins filters $BVI$,\footnote{As in previous works, e.g. \citet{DH05}, the $R$-band filter was not included as it can be affected by H$\alpha$.} and convert to monochromatic fluxes using zeropoints from \citet{Fukugita95}.
 \item Construct SEDs for filter combinations $BV$, $BVI$, $VI$, and model\footnote{All modelling was performed using \textsc{python}.} with a Planck function to find $R_{\rm BB}$ for each SED.
 \item Model time-series spectra to determine the blueshifted velocity of either the Si~\textsc{ii} $\lambda$6355 of Fe~\textsc{ii} $\lambda$5169 transition (Fig. \ref{fig:vels}), and hence calculate $R_{\rm phot} = v_{\rm ph}t$.
 \item Determine $\zeta$ as function of colour temperature.
\end{enumerate}

For the GRB-SNe in the sample, the same photometric/spectroscopic observations are required, but the GRB-SN LCs need to be further decomposed in order to isolate the flux coming from the SN itself.  For every cosmological GRB-SN event, light arises from three sources \citep{Zeh04,Ferrero06,Cano11a,Hjorth13}: the afterglow, the accompanying SN, and a constant source of flux from the underlying host galaxy.  In addition to the host-flux removal, the AG contribution was removed by fitting the host-subtracted LCs with an an afterglow model.  The light that powers a GRB afterglow is expected to be synchrotron in origin, and thus has a power-law dependence in both time and frequency\footnote{$f_{\nu} \propto (t - t_{0})^{-\alpha}\nu^{-\beta}$, where $t_{0}$ is the time at which the GRB triggered by a GRB satellite, and the temporal decay and energy spectral indices are $\alpha$ and $\beta$, respectively.}.  I followed the procedure described in my previous works \citep{Cano13,Cano14,Canoetal14,Cano15,Cano17} to model, and then subtract away the afterglow contribution, thus leaving flux from only the SN.  Once these steps have been taken, observer-frame SEDs were created, I then interpolated to $BVI$~$(1+z)$ wavelengths and extracted the flux.  This SED-interpolation method performs the necessary K-correction so that we examine precise rest-frame $BVI$ filters, thus allowing a direct comparison with the observations of the calibrating SNe, which have also been K-corrected using the SED-interpolation method.

The observations/measurements used in this work are presented in the following figures: the line velocities relative to the explosion epoch are given in Fig. \ref{fig:vels}, while the colour temperatures and BB radii, also relative to the explosion date are given in Fig.~\ref{fig:BB}.  As the ultimate aim of the two-part analysis is the derivation of EPM distances to GRB-SNe, I have also presented the empirically derived dilution factors as a function of temperature in Fig~\ref{fig:DF_all}, and given their values in Table~\ref{table:DF}.

\section{Discussion I: Inter-comparisons between the SE-SN subtypes}
\label{sec:discussionI_intercompare}

Figs.~\ref{fig:BB} and \ref{fig:DF_all} display some interesting behaviour between the different SE-SN subtypes.  First, let us inspect the BB properties in the former, aforementioned figure.  In terms of the colour ($BV$, $BVI$, $VI$) temperatures, At early phases in both $BV$ and $BVI$ there is a lot of scatter, of order $\sim10,000-15,000$~K.  For $BV$, after about 30~days, the temperatures converge towards a common value of $\sim5000-7000$~K.  In $BVI$ there is increased scatter at all phases, and while the colour temperatures tend to cluster for a given subtype, there appears to be more scatter in the SNe Ib sample than the others.  For the $VI$ filter combination, less scatter is seen relative to the other two filter combinations, although at phases less than ten days post explosion, the temperatures range from roughly $10,000-40,000$~K.  However, the scatter reduces dramatically by $25-30$~days, and cluster around a value of $5500-6500$~K.

More variation is seen in the BB radii for each filter combination, where apart from the two SNe IcBL, considerable scatter is seen at all epochs.  The SN Ic subtype is a good example of this, where it is seen that the BB radii of SN~2011bm is the largest of all in the complete sample, while SN~1994I appears to have the smallest radii of the sample.  At early phases, up to about 30~days post explosion, in $BV$ and $BVI$, the SNe Ic and IcBL appear to have larger BB radii than the helium-rich subtypes, though this comparison somewhat vanishes for the $VI$ filter combination.  Considerable scatter is seen in the $BVI$, limiting the amount of conclusions we can draw from a visual comparison.

Next, in all three colours the dilution factors of the SNe Ic, excluding the SNe IcBL, are generally larger than the other subtypes, though there is considerable overlap with all the subtypes.  Interestingly, the dilution factors of the two SN~IcBL appears to be intermediate between the larger SNe Ic values and the smaller SNe Ib values.  SNe IIb appear to have the largest amount of scatter of all the subtypes, where they overlap with with the larger SNe Ic values, but also extend down to the smallest values in all three filter combinations.  Interestingly, in $BV$ the SN Ib values cluster very closely, possibly suggesting similar physical conditions being present in these events.  Moreover, in $VI$, the SNe Ic/IcBL also cluster relatively tightly at larger temperatures, while just the SNe Ic cluster reasonably well at all $VI$ colour temperatures.  This latter observation hints that SNe Ic over a certain time range (e.g. \citealt{Dessart15}) or temperature range have the potential to be used as a calibrator for GRB-SNe.  This will be explored in more detail in Paper II.

That less scatter is generally seen in dilution factors derived from the $VI$ BB temperatures can be directly related back to the physical conditions of the SN atmospheres.  At bluer wavelengths, the optical magnitudes, especially in $U$ and $B$, are strongly affected by the metal content of a given SN, where increased metallicity leads to increased amounts of line blanketing by iron-peak elements in the ejecta, especially Fe~\textsc{ii}, Cr~\textsc{ii} and Ti~\textsc{ii}. In contrast, at redder wavelengths/filters, especially $V$ and $I$ (excluding $R$ which is affected by H$\alpha$), the photometry is less affected by the metal content, and therefore provides a better estimate of the evolution of the photosphere.

\section{Discussion II: Comparison with the radiative transfer models of D15}
\label{sec:discussionII_RT_compare}

\begin{figure*}
 \centering
 \includegraphics[width=\hsize]{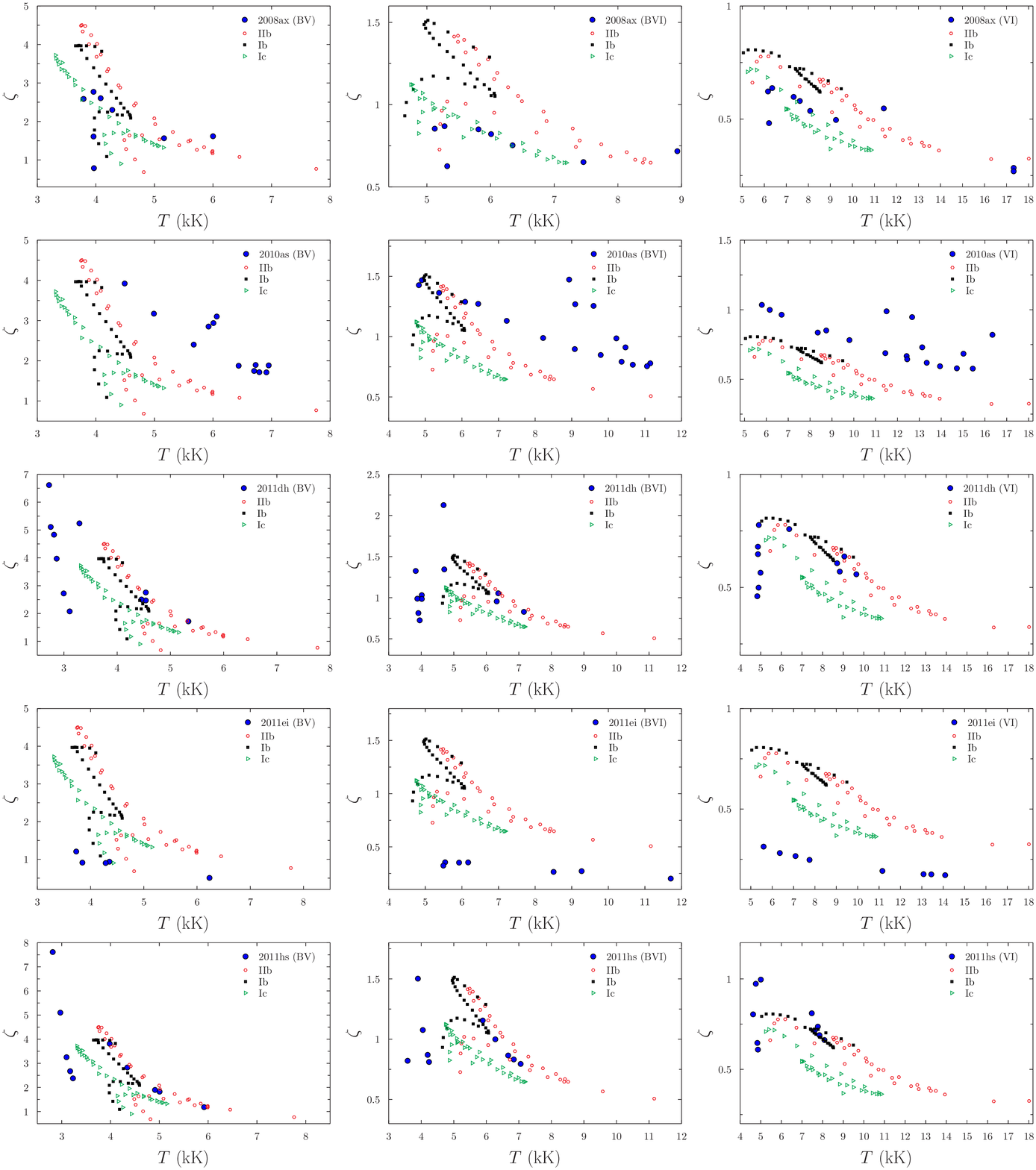}
 \caption{Empirically derived dilution factors of SNe IIb 2008ax, 2010as, 2011dh, 2011ei and 2011hs.  Plotted for comparison are synthetic dilution factors for SNe IIb (red), Ib (blue) and Ic (green) from D15 $-$ see Sect. \ref{sec:discussionII_RT_compare} for an explanation of the model inputs.}
 \label{fig:IIb_vs_D15}
\end{figure*}

\begin{figure*}
 \centering
 \includegraphics[width=\hsize]{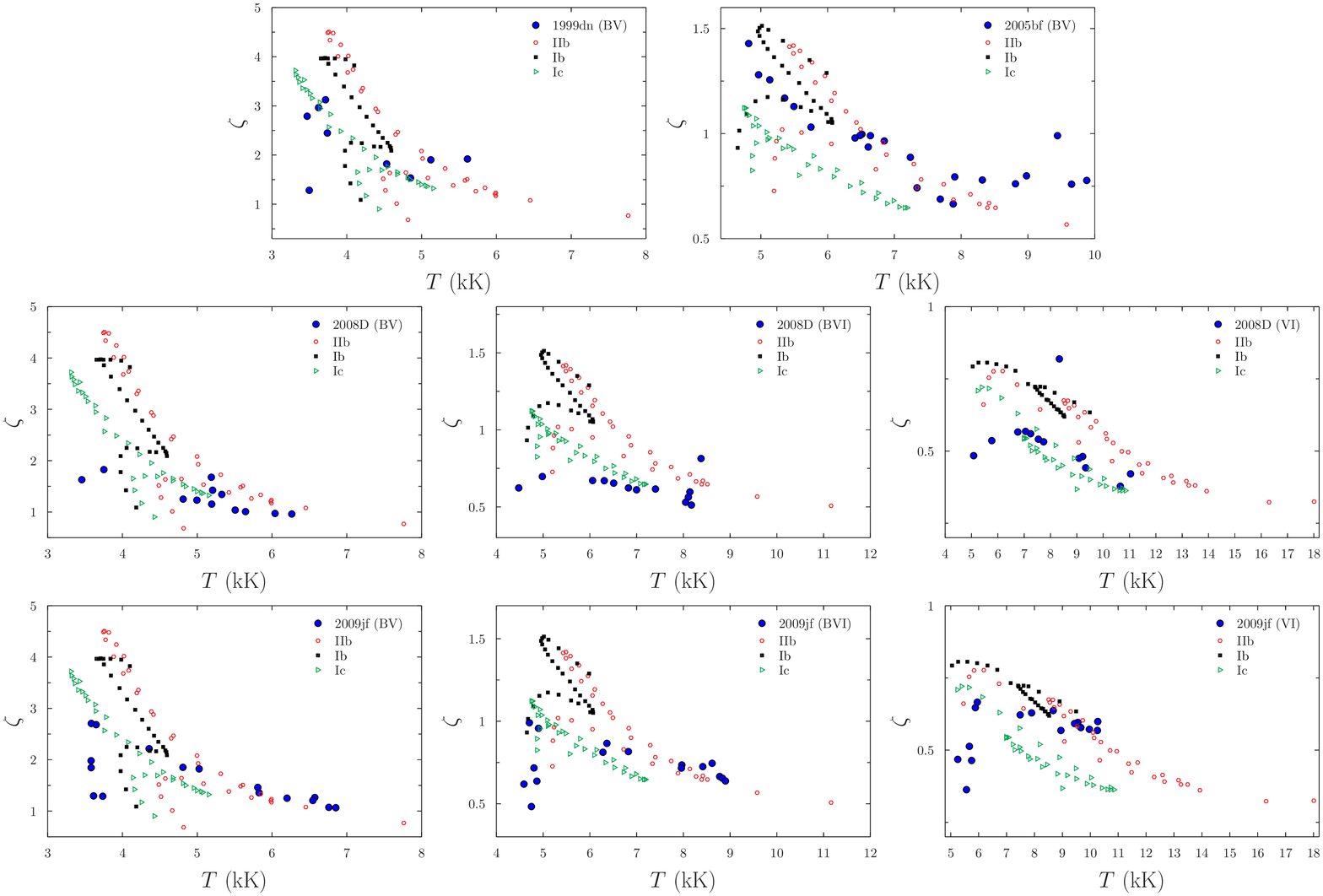} 
 \caption{Empirically derived dilution factors of SNe Ib 1999dn, 2005bf, 2008D and 2009jf.  Plotted for comparison are synthetic dilution factors for SNe IIb (red), Ib (blue) and Ic (green) from D15 $-$ see Sect. \ref{sec:discussionII_RT_compare} for an explanation of the model inputs.}
 \label{fig:Ib_vs_D15}
\end{figure*}

\begin{figure*}
 \centering
 \includegraphics[width=\hsize]{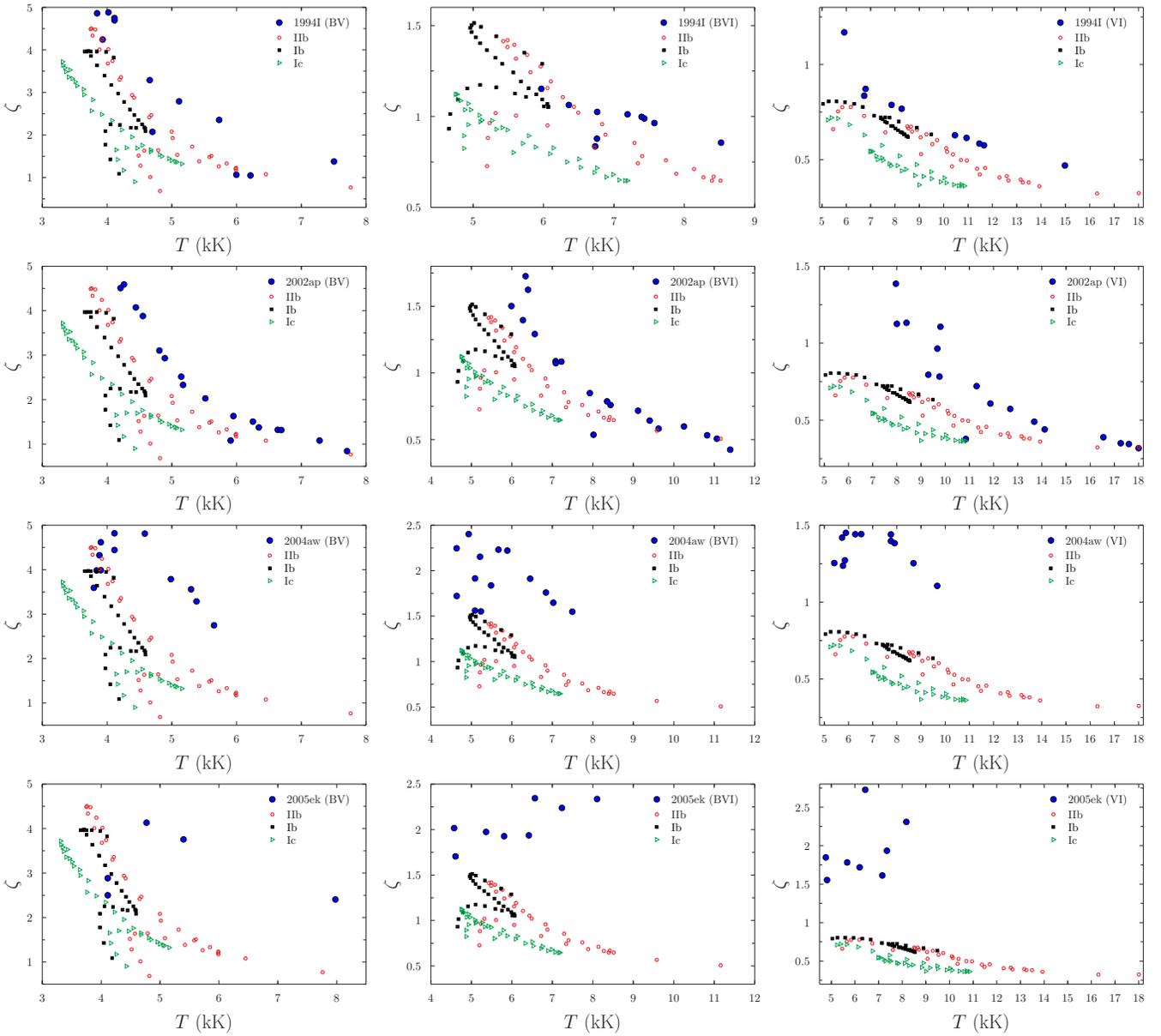} 
 \caption{Empirically derived dilution factors of SNe Ic 1994I, 2002ap, 2004aw and 2005ek.  Plotted for comparison are synthetic dilution factors for SNe IIb (red), Ib (blue) and Ic (green) from D15 $-$ see Sect. \ref{sec:discussionII_RT_compare} for an explanation of the model inputs.}
 \label{fig:Ic_vs_D15_1of2}
\end{figure*}

\begin{figure*}
 \centering
 \includegraphics[width=\hsize]{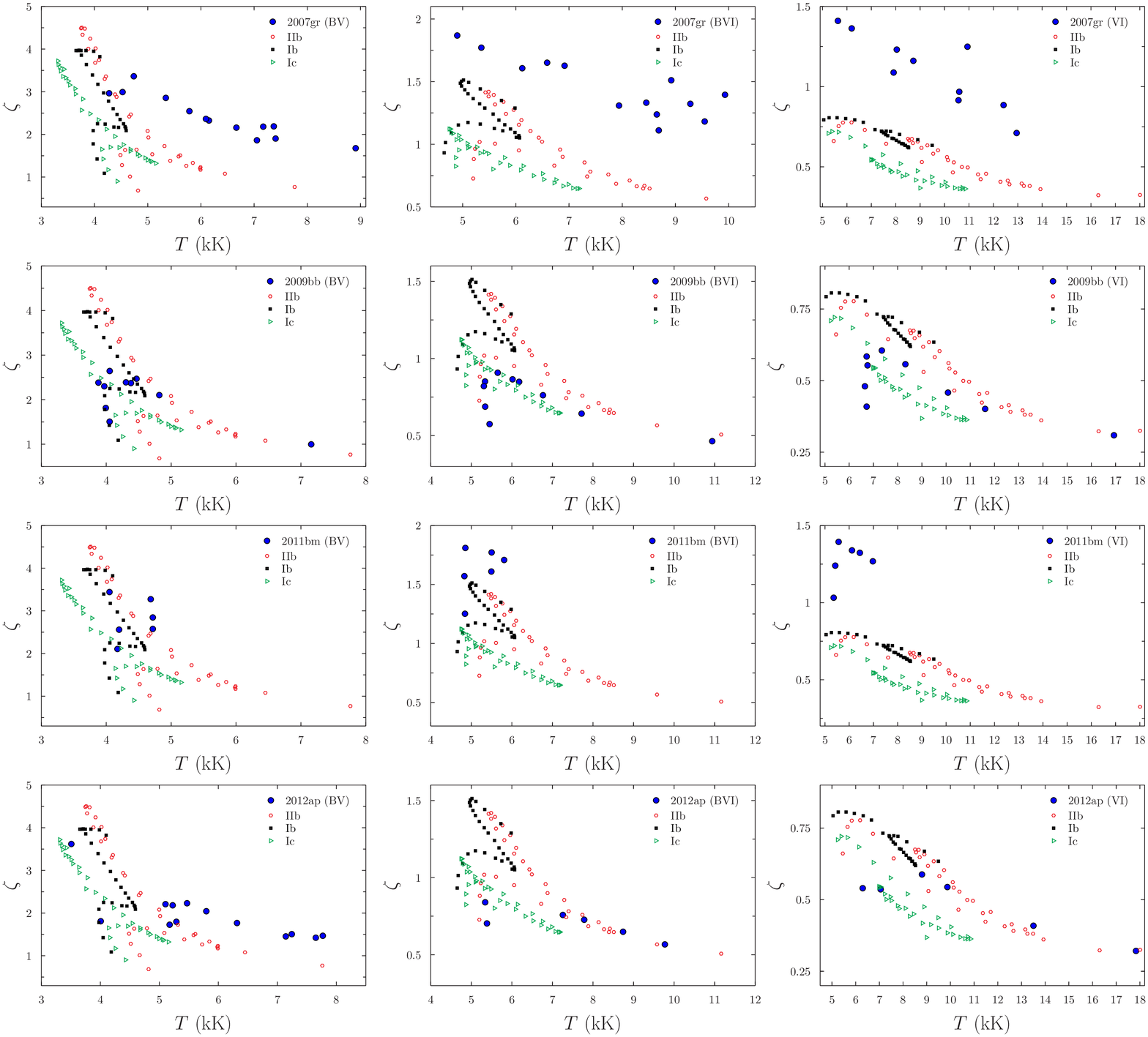} 
 \caption{Empirically derived dilution factors of SNe Ic 2007gr and 2011bm, and the two relativistic SNe IcBL 2009bb and 2012ap.  Plotted for comparison are synthetic dilution factors for SNe IIb (red), Ib (blue) and Ic (green) from D15 $-$ see Sect. \ref{sec:discussionII_RT_compare} for an explanation of the model inputs.  It is seen that under the assumption that all of the SNe Ic considered here arise from binary systems, only the relativistic SNe IcBL (2009bb and 2012ap) are consistent with the SNe Ic models of D15, while all others are more consistent with the SNe IIb models.}
 \label{fig:Ic_vs_D15_2of2}
\end{figure*}

Figures \ref{fig:IIb_vs_D15}, \ref{fig:Ib_vs_D15}, \ref{fig:Ic_vs_D15_1of2} and \ref{fig:Ic_vs_D15_2of2} show a comparison of the empirically derived dilution factors of the six SNe Ic and two SNe IcBL in the sample with those calculated from the non-local thermodynamic-equilibrium RT models from D15.  The RT models of D15 consider the terminal explosion of the primary star in a close-binary system.  The RT simulations are time-dependent, and include both non-thermal and non-local energy deposition effects, and different amounts of chemical mixing in the SN ejecta.  The progenitor models used in the RT simulations are taken from \citet{Yoon10}, which simulated the evolution of a close-binary system from the main sequence to the onset of Ne burning.  The hydrodynamical models of \citet{Yoon10} include differential rotation, tides, mass and angular momentum transfer, and stars of mass 12$-$60~M$_{\odot}$ and mass ratios of $\le1-1.5$.  Later evolution and the subsequent explosion of the primary was performed using \textsc{kepler} \citep{Weaver78}, where the explosion is artificially conducted using a piston.


The models considered in D15 are listed below, where each is a representation of a possible progenitor star system of a SN IIb (D15: 3p65Ax1), SN Ib (D15: 6p5Ax1) and SN Ic (D15: 5p11Ax1).  

\begin{itemize}
 \item \textbf{IIb model}: 
  \subitem $-$ Initial mass of 16~M$_{\odot}$, final mass of 3.65~M$_{\odot}$, radius of 1.2 solar radii and a temperature of 21,000~K.
  \subitem $-$ SN kinetic energy of $1.24 \times 10^{51}$~erg, ejecta mass of 2.22~M$_{\odot}$ and remnant mass of 1.43~M$_{\odot}$.
  \subitem $-$ The exploding star has a sizable He-rich shell and some residual H (0.005~M$_{\odot}$) in the outermost layers.  Representative of either a WN or WNh type star.
 \item  \textbf{Ib model}: 
  \subitem $-$ Initial mass of 25~M$_{\odot}$, final mass of 6.5~M$_{\odot}$, radius of 2.0 solar radii and a temperature of 68,700~K.
  \subitem $-$ SN kinetic energy of $1.26 \times 10^{51}$~erg, ejecta mass of 4.97~M$_{\odot}$ and remnant mass of 1.53~M$_{\odot}$.
  \subitem $-$ Contains 1.65~M$_{\odot}$ of He (35\% of the total mass), and no H.  The explosion produces He lines in the synthetic spectra.  Representative of either a WN or WNh type star.
 \item \textbf{Ic model}: 
  \subitem $-$ Initial mass of 60~M$_{\odot}$, final mass of 5.11~M$_{\odot}$, radius of 5.2 solar radii and a temperature of 130,000~K.
  \subitem $-$ SN kinetic energy of $1.25 \times 10^{51}$~erg, ejecta mass of 3.54~M$_{\odot}$ and remnant mass of 1.57~M$_{\odot}$.
  \subitem $-$ Typical composition of a WC star.  He, C and O are in the outermost layers. Even with strong mixing no He lines are present in the synthetic spectra.
\end{itemize}

The kinetic energy of each explosion is essentially the same (roughly 1.2 Bethe), while the remnant masses are in the range 1.4$-$1.5~M$_{\odot}$.  The ejecta masses range from 3$-$5~M$_{\odot}$.  The initial masses increase from IIb (16~M$_{\odot}$) to Ib (25~M$_{\odot}$) to Ic (60~M$_{\odot}$).  It is also seen that the larger the initial mass, the more extended the star is.  There is no correlation of the final masses or ejecta masses with the initial masses.  Finally, at a given temperature, dilution factors arising from less massive progenitors will generally be larger than those arising from more massive progenitors.

The following subsections provide an in-depth look and discussion of each SE-SN subtype relative to the D15 models.  One caveat to be mindful of is that the following comparisons are made under the assumption that all of the SE-SNe here arise from close-binary systems (as the D15 synthetic values have been derived for this scenario).

\subsection{SNe IIb}

\begin{itemize}
 \item \textbf{2008ax}: In all three filter combinations, the empirical dilution factors are consistent with the green (SN Ic) points.  In $BV$ there is also overlap with the black (SN Ib) points and with the red (SN IIb) points at temperatures greater than 5000~K.  There is very little/no overlap with the SN IIb and Ib synthetic dilution factors in $BVI$ and $VI$.  Interestingly, the sharp drop-off at lower temperatures seen in the D15 models are also seen in the empirical values.
 \item \textbf{2010as}: This is the only SN IIb in the sample where the empirical dilution factors sit above the synthetic points in all three filter combinations, and is hence inconsistent with all of the D15 models.
 \item \textbf{2011dh}: The empirical dilution factors probe a region of the diagrams not populated with values from the D15 models. In regions where there is overlap, generally there is reasonable agreement with the SN IIb and Ib models.  As seen also for SN~2008ax, there is a sharp turnoff at lower temperatures, but in this case the turnover occurs at lower temperatures than seen in the D15 models.
 \item \textbf{2011ei}: This is another interesting event whereby the empirical dilution factors are all lower than each D15 model, and there is essentially no overlap between them. Of all the SNe IIb investigated here, SN~2011ei has the smallest dilution factors.
 \item \textbf{2011hs}: At lower temperatures, the empirical dilution factors probe regions of the diagram not populated by the D15 models.  In regions where they do overlap, there is good agreement with the IIb model in $BV$ for temperatures greater than 4000~K.  The same also applies for $BVI$ for temperatures greater than 6000~K.
\end{itemize}

In the small sample investigated here, there are no general trends in the empirical SN IIb dilution factors. It could be argued that SN~2008ax is most consistent with the D15 SN Ic values, while only SN~2011dh, and to a lesser extent SN~2011hs are loosely consistent with the synthetic SN IIb points. Neither SN~2010as nor SN~2011ei show much agreement with any of the D15 models.

\subsection{SNe Ib}

\begin{itemize}
 \item \textbf{1999dn}: Only empirical dilution factors for the filter combination $BV$ are presented here.  There is a lot of scatter in their values, however there is a hint of more agreement with the D15 SN Ic model than the other two synthetic values.  There is also a hint of a turnover at lower temperatures, but the amount of scatter present limits how certain we can conclude this.
 \item \textbf{2005bf}: As for SN~1999dn, only the $BV$ filter combination is investigated here. The empirical dilution factors occupy regions above the SN Ic synthetic values but below the SN IIb values.  While there is not perfect agreement with the D15 SN Ib values, they are clearly inconsistent with the SN IIb and Ic values.
 \item \textbf{2008D}: Of the four SNe Ib investigated here, SN~2008D has the smallest dilution factors. In all three filter combinations there is a hint of a turnover at lower temperatures, however its evolution is unlike all the other SE-SNe in the same, which is much smoother. In all three colours, the empirical values either occur below all three models (i.e. $BVI$), or they overlap with the SN Ic D15 values (this is particularly so for $VI$).
 \item \textbf{2009jf}: For this event, there is a much more apparent, sharp turnover at lower temperatures, where especially in the $BVI$ filter combination, the turnover occurs at approximately the same temperature as the SN Ib and Ic D15 values. In all three colours there is a lot of overlap with the synthetic SN IIb values at larger temperatures, while at lower temperatures there appears to be more agreement with the SN Ic values.
 \end{itemize}

In the four-object sample of SNe Ib presented here, there is a small suggestion of agreement between the SN Ib and Ic synthetic dilution factors and the empirically derived ones. One possible exception is for SN~2005bf, for which one could argue for agreement with the SN IIb D15 values. The most extreme values were seen for SN~2008D, where the empirical values were the smallest in the sample, and generally lower than the D15 models. These extreme values appear to be similar to those measured for the two relativistic SNe IcBL (see the following section), but they evolve differently at lower temperatures, where those of SN~2008D evolve more smoothly and do not have a sharp turnoff.

\subsection{SNe Ic \& IcBL}

\begin{itemize}
 \item \textbf{1994I}: In all three filter combinations, the empirical dilution factors are either above ($BV$ and $VI$) or overlap ($BVI$) with the SN IIb D15 points.
 \item \textbf{2002ap}: In all three filter combinations, the empirical dilution factors are all above the SN IIb D15 values, with a slight hint of convergence with the SN IIb values at large temperatures.
 \item \textbf{2004aw}: In $BVI$ and $VI$,  the empirical dilution factors are well above the synthetic SN IIb points.  In $BV$ there is overlap with both the SN IIb and SN Ib points around 4000~K.
 \item \textbf{2005ek}: Exactly the same as SN~2004aw.
 \item \textbf{2007gr}: The same as SNe 2004aw and 2005ek, but overlapping values around 4500~K.
 \item \textbf{2009bb}: In all three filter combinations, the empirical dilution factors mostly overlap with the SN Ic D15 points, although there is also overlap with the SN Ib points in $BV$.
 \item \textbf{2011bm}: In $BV$ and $BVI$, the empirical dilution factors are consistent with the SN IIb and SN Ib D15 values at lower temperatures ($\sim4800$~K), and in $VI$ they are always above the SN IIb points.
 \item \textbf{2012ap}: In $BV$ and $BVI$ there is overlap with the SN Ic and possibly the SN IIb D15 points, where there is excellent overlap with the SN IIb values in $BVI$ at temperatures above 7500~K.  In $VI$ there is overlap with all three synthetic datapoints, and at temperatures in excess of 13,500~K, they match the SN IIb points very well.
\end{itemize}


Of the six SNe Ic and two relativistic SNe IcBL studied here, only the latter are consistent with the SNe Ic progenitors in the RT models of D15.  All of the remaining SNe Ic occupy the regions consistent with, or above, the regions populated by the synthetic SNe IIb values.

\section{Discussion III: Comparison with GRB-SNe}
\label{sec:discussionIII_GRBSNe}

\begin{figure*}
 \centering
 \includegraphics[width=\hsize]{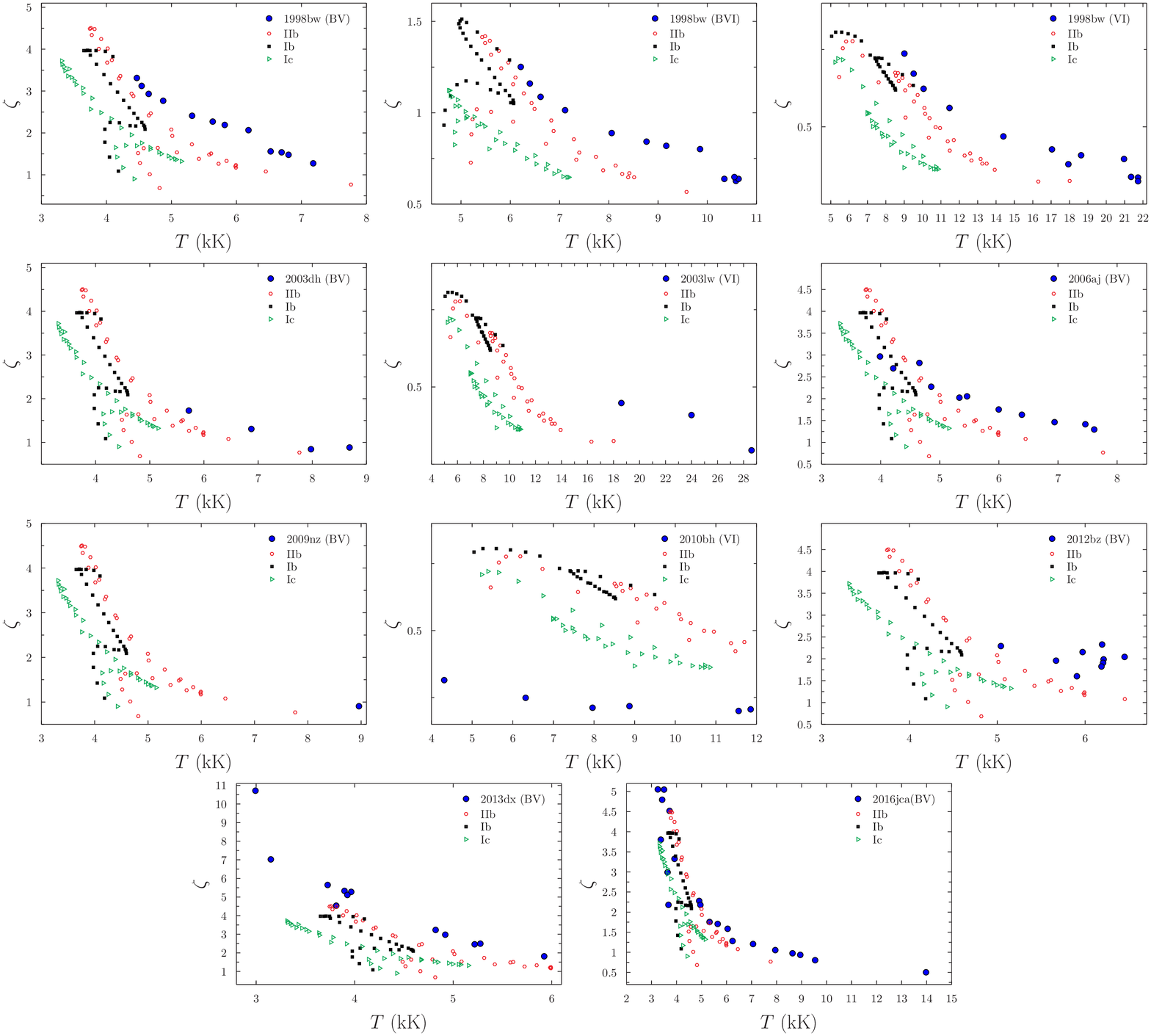} 
 \caption{Dilution factors of GRB-SNe, calculated using luminosity distances for a generic, flat $\Lambda$CDM cosmology (see Section~\ref{sec:discussionIII_GRBSNe}).  It can be seen that, particularly for SNe~1998bw and 2016jca, the dilution factors occupy an area of the diagram also occupied by SNe 1994I and 2002ap, which might hint at the suitability of the latter as calibrators (see Paper II).  A clear exception is SN~2010bh, whose lower dilution factors are a direct result of its higher velocity Si \textsc{ii} $\lambda$6355 velocities relative to the other GRB-SNe (Fig. \ref{fig:vels}).  None of the GRB-SNe investigated here have the same dilution factor as a function of temperature as that measured for the two relativistic SNe IcBL 2009bb and 2012ap, indicating that their stellar progenitors may be fundamentally different.  }
 \label{fig:GRB_vs_D15}
\end{figure*}

We saw in the previous section that under the assumption that the SNe Ic in the sample arise from close binary systems, only SNe~2009bb and 2012ap are consistent with the SNe Ic models of D15.  In turn, if GRB-SNe and SNe IcBL arise from similar types of progenitors, as suggested by several authors \citep{Cano13,Modjaz16,CanoReview17}, then as a first-order approximation, one might anticipate that GRB-SNe will have dilution factors that are similar to those of relativistic SNe IcBL as this would reflect that both sub-types will have similar atmospheres.  As a preliminary comparison, plotted in Fig. \ref{fig:GRB_vs_D15} are the dilution factors of GRB-SNe, as calculated using the distance luminosity for a generic, flat $\Lambda$CDM cosmology ($H_{0} = 70$~km~s$^{-1}$~Mpc$^{-1}$, $\Omega_{\rm M} = 0.7$ and $\Omega_{\rm \Lambda} = 0.3$).  It is important to recall that all the dilution factors of the SE-SNe have been calculated using model-free distances.  In the following I more carefully digest the empirical dilution factors of the GRB-SN sample relative to the D15 synthetic values.

\begin{itemize}
 \item \textbf{1998bw}: Due to the close proximity of this event relative to all other GRB-SNe, high-cadence photometric and spectroscopic observations were obtained. This has allowed us to investigate all three filter combinations for this single GRB-SN. In general, as was also observed for some of the SNe Ic, in particular SNe~1994I and 2002ap, the empirical dilution factors all occur above the synthetic values from the D15 models. 
 \item \textbf{2003dh}: Only $BV$ temperatures greater than $\sim6000$~K are probed by the observations, which are generally larger than the synthetic SN IIb values.
 \item \textbf{2003lw}: Similar to SN~2003dh, the $VI$ colour temperatures are above those of the SN IIb, but there is very little overlap with the synthetic values.
 \item \textbf{2006aj}: At $BV$ colour temperatures less than 5000~K, there is some overlap with the SN IIb and Ib values, while at larger temperatures the synthetic values are all above the SN IIb D15 points.
 \item \textbf{2009nz}: The solitary empirical dilution factor in $BV$ occurs at too large a temperature ($\sim9000$~K) where there are no D15 values to compare with.
 \item \textbf{2010bh}: Very interestingly, this event displays the most extreme (low) empirical dilution factors of the GRB-SNe studied here. The $VI$ values are all lower than the D15 points.  This is likely because of the large inferred photospheric velocities (Fig.~\ref{fig:vels}), and hence large photospheric radii and smaller dilution factors.
 \item \textbf{2012bz}: The $BV$ colour temperatures probed here are all greater than 5000~K, where the empirically derived dilution factors are all above the synthetic SN IIb points.
 \item \textbf{2013dx}: As for the majority of the other GRB-SNe, the empirical values are all above the SN IIb D15 values.
 \item \textbf{2016jca}: This event is the only where there is a hint of a turnover at lower temperatures.  Indeed, of all the GRB-SNe investigated here, the $BV$ colour temperatures extend to the smallest values. At temperatures larger than the turnover temperature (the latter around $\sim3500$~K), the points are just above, or vaguely coincide with the SN IIb values.  The turnover values probe regions occupied by the synthetic SN IIb and Ic values.
\end{itemize}

Interestingly, the dilution factors of all GRB-SNe except for SN~2010bh are either consistent with, or occupy spaces in the plot above the red (SN IIb) points.  This is entirely inconsistent with the regions populated by the empirical dilution factors of the two relativistic SNe IcBL.  Instead, it is seen that the GRB-SN dilution factors occupy an area of the diagram also occupied by SNe 1994I and 2002ap, which may hint at the latter as suitable proxies for the dilution factor vs. temperature of GRB-SNe. This will be investigated in detail in Paper II.   An exception however is SN~2010bh, whose lower dilution factors are a direct result of the higher velocity Si \textsc{ii} $\lambda$6355 velocities relative to the other GRB-SNe (Fig. \ref{fig:vels}).  Only SN~2016jca probes $BV$ colour temperatures low enough that a turn over is seen, which occurs around $\sim3500$~K, which is roughly the turnover temperature seen for the synthetic SN Ic values.

Of great interest to our story is the comparison of the GRB-SNe dilution factors with those of the two relativistic SN IcBL.  The presence of a central engine has been inferred for both types of SNe, which can either be a rotating black hole surrounded by an accretion disk, or a rapidly rotating neutron star.  The key difference between them however is that in the latter events, it is thought that the jet produced by the central engine does not escape all the way into space, thus explaining why $\gamma$-ray emission was not detected for SNe~2009bb and 2012ap.  Two explanations have been proposed for this: (1) either the central engines of SNe IcBL are less energetic than those that produce a GRB, and the jet is not supported long enough for it to break out of the star; or (2) the progenitors of SNe IcBL are more extended than those of GRB-SNe.  Of course both factors could be acting simultaneously.  

Taking the results here at face value (and assuming that GRB-SNe and relativistic SNe IcBL both arise from binary systems), it appears that the progenitors of the two relativistic SNe IcBL are more massive, and more extended, than the progenitor stars of the GRB-SNe in the sample.  It is possible to draw this conclusion based on the properties of the D15 models, where the progenitor in the SN Ic model (of which the empirical dilution factors of the two relativistic SNe IcBL most closely resembled) is more massive and more extended than the model SN IIb progenitor (of which the empirical GRB-SNe dilution factors more closely resembled).  This then suggests that the reason a GRB was not detected in the former is because the stellar progenitors are too fat to be penetrated by the jet.  Of course one cannot rule out however than a less energetic engine is also at play, or combinations of both effects.  Secondly, that the GRB-SNe arise from less massive progenitors may also be hinting at the physical processes that produce a GRB in the first place: if the progenitors of GRB-SNe are actually binary systems, then interactions between the two can ultimately produce the $\gamma$-ray emission.  Such a scenario can arise if they share a common envelope phase, and then the inspiral of a compact object arising from the SN of one component into the core of the other, which spins up the secondary's core, providing the necessary angular momentum to power the central engine.  In addition, binary interactions have stripped much of the outer layers into space, so that once the jet is formed, it must only bore through a few solar radii of material before escaping into space.  All of these factors then combine to create the conditions necessary to produce a GRB.  

However, the fact that only the empirical dilution factors of SNe~2009bb and 2012ap match those derived from the RT modelling of D15, while those of all others do not, may indicate that the latter do not arise from binary systems.  With mounting evidence for the preferential occurrence of massive stars in binary systems (e.g. \citealt{Sana2012}), opinion is growing that the binary star evolution scenario may be the dominant channel for SNe Ibc production (e.g. \citealt{Podsiadlowski92,Yoon15}).  However, the results here indicate that if SNe Ic arise from binary systems, there are missing ingredients in the advanced modelling studies of D15 that must be responsible for the discrepancy between the observations and models.  Instead, the results here still support the suggestion that most SNe Ic, including GRB-SNe, arise from single-star progenitors, which may have similar ranges of initial masses.  For example, the results of \citet{Fruchter06} and \citet{Kelly08}, which showed that the similar distribution of GRB-SNe and SNe Ic in their host galaxies implies comparative ranges of progenitor masses.  

\section{Considerations \& Caveats}
\label{sec:caveats}

\subsection{The Expanding Photosphere Method}
\label{sec:Caveats}

As noted by other authors (e.g. \citealt{Vinko2004,DH05}), the main assumptions of the EPM are:

\begin{enumerate}
 \item The SN is spherically symmetric, and the ejecta is homologous.
 \item There is a unique and well-defined photospheric radius, where the ejecta below this radius is optically thick.
 \item The photosphere radiates as a diluted (parametrized by the dilution factor) blackbody.
\end{enumerate}

Starting with assumption (i), it has been shown that SNe Ic such as SN~2002ap, and GRB-SNe, are expected to possess different degrees of asphericity (see \citealt{WangWheeler08,CanoReview17} for extensive reviews of the geometry of core-collapse SNe and GRB-SNe, respectively).  As a relevant example for the discussion here, considerable attention has been given to the geometry of SN~1998bw, where its aspherical, axis-symmetric geometry has been ascertained, either directly from observations (e.g. polarimetry: \citealt{Patat01}) or from detailed modelling of photospheric and nebular spectra.  \citet{Maeda06} demonstrated via 3D RT simulations that there was a modest amount of boosted luminosity emanating the polar axis relative to the equatorial axis.  This ratio decreased as SN~1998bw approached maximum light, reaching a factor of $\sim1.2$ until the nebular phase (+60 d; \citealt{Patat01,Mazzali01}).  Other works by \cite{Maeda02,Nakamura01,Hoeflich99} and \citet{Mazzali01} also concur with the notion that SN~1998bw has an asymmetric SN ejecta structure.  Additionally, spectropolarimetric observations were obtained of SN~2002ap \citep{Kawabata02,Leonard02, Wang03}, and polarization at the 1--2\% level was observed in all three papers.  Each group concluded on its aspherical nature; for example, \citet{Leonard02} suggested that SN~2002ap possessed an asymmetry of 15$-$20\% during its photospheric and early nebular phases, as well as a complex morphology of the thinning ejecta.  

It was suggested by \citet{Cano14} that most GRB-SNe are seen at, or quite close to, the same viewing angle in each event $-$ i.e. close to the jet angle.  This provides one possible reason as to why GRB-SNe are standardizable candles.  It was also seen that relativistic SNe IcBL 2009bb and 2012ap followed the same luminosity$-$stretch \citep{Cano14} and LDR \citep{CJG14,LiHjorth14} as observed for GRB-SNe.  This was attributed to SNe~2009bb and 2012ap being observed at similar viewing angles as GRB-SNe.  However, SN~2002ap did not follow these relations, meaning that the elongation axis in this event was likely pointed away from earth.  Moreover, the fact that SN~2002ap was not associated with a GRB event \citep{Hurley02}, led \citet{Mazzali2002} to conclude that the level of asphericity in SN~2002ap is not as severe as that expected for GRB-SNe.


Next, assumption (ii) is only valid at early times when the outflow is fully ionized, and effects due to line blanketing are small \citep{DH05}.  This applies to photospheric epochs only, and the method clearly breaks down as the SN transitions into nebular phases. This assumption has received an lot of consideration over the years (e.g. \citealt{DH05,Dessart15}), and I encourage the reader to consult these papers for further discussion.  For the sake of this study, it is worth remembering that the definition of a discrete photosphere does not exist in nature.  Instead, at least during the photospheric phase, it is more precise to consider a photospheric \emph{region}.  This arises from the fact that the optical opacity is strongly wavelength-dependent, and in the rapidly expanding SN ejecta, the photosphere does not have a unique, well-defined spatial location.  Even photons with identical wavelengths will escape not at a specific radius, but rather over a region.  Blue and UV photons will suffer more line-blanketing effects than redder photons,  which leads to a larger effective escape radius for the former.  All of these effects can lead to decoupling radii for photons of similar wavelengths that vary by factors of 2$-$3 (D15). 

In this work, I was forced to use the approximation that the photosphere has a distinct location in space.  Its precise value has been inferred using the blueshifted velocity of Si \textsc{ii} $\lambda$6355 or Fe~\textsc{ii}~$\lambda5169$.  Discussion presented in the literature has suggested that using the blueshifted velocity of specific line transitions as a proxy for the photospheric velocity is probably reasonable as a first-order approximation (e.g. \citealt{Valenti2012,Modjaz16}).  I have attempted to eliminate some uncertainty in the analysis by using the same line transition as a proxy for each SN in the hope that the dilution factors derived using this crude approximation are affected the same way for each SN considered here.

\subsection{Error propagation}

All observational errors and those from the linear distances (Table \ref{table:vitals}) have been propagated through to the derived dilution factors and GRB distances.  The main sources of error arise mostly from the published observational data, with the dominant being the uncertainties in the line velocities velocities and the distances to the SNe host galaxies.  For all cases, the errors in $\xi$ and $D$ were determined via the quotient rule: $\Delta x$~$=$~$ x \sqrt{(\frac{\Delta y}{y})^{2} + (\frac{\Delta z}{z})^{2}}$.

\section{Conclusions \& Future Prospects}
\label{sec:conclusions}

In this work, and in the context of the EPM, the empirical dilution factors of a sample of SE-SNe, including SNe IIb, Ib, Ic, relativistic IcBL and GRB-SNe, have been presented.  The ultimate goal of this project is to use the empirically derived dilution factors of either the entire SE-SN sample, or perhaps just the SNe Ic/IcBL, as direct proxies for the dilution factors of GRB-SNe, and hence facilitate their use as cosmological probes.  In this paper, which is the first of a two-part series, I presented my method for obtaining the empirical dilution factors from photometric and spectroscopic observations of the SNe themselves.  The SE-SNe sample was chosen on the basis that cosmological-model-independent distances exist to their host galaxies, and each had a sufficient dataset that allowed for host-subtracted, dereddened rest-frame $BVI$ LCs to be constructed, time-series spectra, and knowledge of the entire line-of-sight extinction.  The photometric data allowed for us to model host-subtracted, dereddened SEDs with a Planck function to derive the blackbody radius and blackbody temperature as a function of time (Fig.~\ref{fig:BB}).  Similarly, the dereddened, host-corrected time-series spectra were used to model and determine the blueshifted velocity of either Si \textsc{ii}~$\lambda$6355 (SNe Ic, IcBL and GRB-SNe) or Fe~\textsc{ii}~$\lambda$5169 (SNe IIb and Ib), which was used a proxy of the photospheric velocity (Fig.~\ref{fig:vels}).  The ratio of these empirically derived radii was taken as the dilution/correct factor, which was plotted as a function of temperature (Figs. \ref{fig:IIb_vs_D15}, \ref{fig:Ib_vs_D15}, \ref{fig:Ic_vs_D15_1of2}, and \ref{fig:Ic_vs_D15_2of2}, see as well Table~\ref{table:DF}).  Dereddened observations of the GRB-SNe were also modelled in the same manner, but instead luminosity distances were used to calculate the empirical dilution factors.

Next, I compared our observationally derived dilution factors with those obtained from the radiative transfer models that were calculated for binary stars (D15).  The small samples investigated here do not allow us to draw definite conclusions regarding the nature of the (possible) binary progenitors of SE-SNe, however a few general trends were seen. First, for the SNe IIb, there was no consensus between the synthetic and empirical dilution factors, where only SN~2011dh, and possibly SN~2011hs were consistent with the SN IIb D15 values.  Indeed it was seen that the empirical dilution factors of SN~2008ax were more consistent with the SN Ic D15 points. For the SNe Ib, there was a hint of agreement between the empirical values and the SN Ib and Ic D15 values. One exception is SN~2008D, which had the most extreme (lowest) values of the SN Ib sample. Moreover, while a sharp turnover was seen at lower temperatures for many of the SE-SNe here (very similar to those seen in the D15 models), only SN~2008D showed a very smooth and gradual turnover, relative to the sharp transition seen both in observations and the D15 models.  

Interestingly, the empirical dilution factors of the SNe Ic and GRB-SNe studied here occupied regions in the diagram populated by the SN IIb D15 values, or above them.  The exception to this were the two relativistic SNe IcBL, which had generally smaller empirical values, which appeared to match those of the SN Ic D15 models.  If we take this result at face value, it hints at differences in the physical properties of the progenitors of relativistic SNe IcBL relative to GRB-SNe.  If both sets of SNe come from binary systems, then those of GRB-SNe are less massive and less extended than those of the relativistic SNe IcBL.  This may be hinting at why high-energy emission is seen for GRB-SNe but not for relativistic SNe IcBL, as the latter are more extended, so that any jet that is produced during the core collapse does not operate long enough for it to escape into space.  However, we cannot rule out a less powerful engine at play in the latter events.  Moreover, the results here may also imply that relativistic SNe IcBL and GRB-SNe have fundamentally different progenitor systems, where the former arise from massive single stars, and the latter from binary systems, where perhaps orbital angular momentum is converted into core angular momentum at the time of merger/collapse, which conspire to produce a central engine and thus a $\gamma$-ray pulse.  This supposition is of course tentative, however we can say with some certainty that the dilution factors between these two types of engine-driven events are quite different, and hint at fundamental differences in their progenitor systems. 

Finally, in Paper II I will use these empirically derived dilution factors to obtain model-free distances to the hot galaxy of one or more GRB-SN, and hence use them as cosmological probes.

\section{Acknowledgments}

I is very grateful for the discussion with Mattias Ergon on deriving the empirical dilution factors of the SNe presented here. I am also very grateful for fruitful discussions with Joszef Vink\'o, Palli Jakobsson, Peter Hoeflich, Steve Schulze and Keiichi Maeda regarding the content of this analysis. Additional gratitude is extended to Elena Pian for making the Si {\sc ii} $\lambda$6355 velocities of SN~1998bw available to me. 

ZC acknowledges support from the Juan de la Cierva Incorporaci\'on fellowship IJCI-2014-21669 and from the Spanish research project AYA 2014-58381-P.

\bibliographystyle{mn2e}

\appendix

\section{Blackbody Fits}

\begin{landscape}
\begin{figure*}
 \centering
 \includegraphics[bb=-52 -34 998 395,keepaspectratio=true, width=\hsize, trim={0 0 0 -200}]{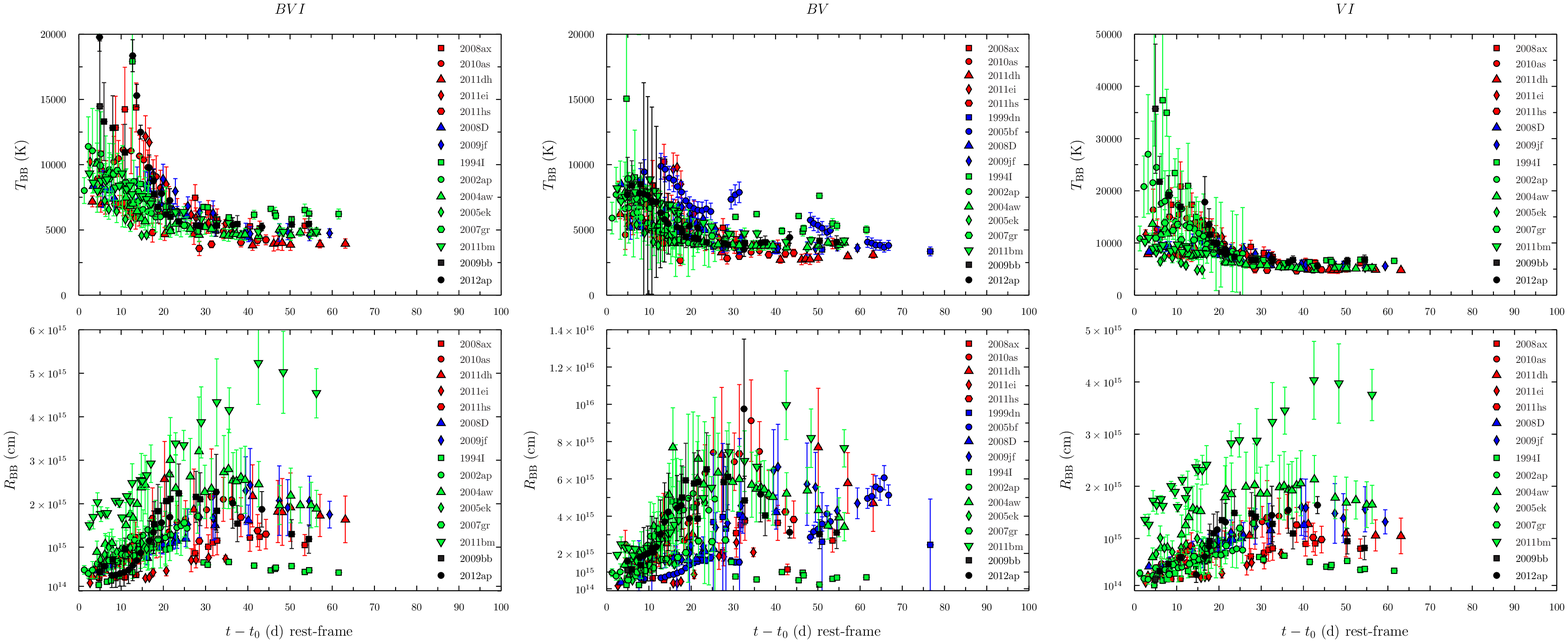} 
 \caption{Blackbody colour temperatures and radii of the SNe IIb (red), SNe Ib (blue), SNe Ic (green) and the two relativistic SNe IcBL (black) in our sample.  Time are given in the rest frame relative to the explosion epoch. \textit{Top}: Colour temperatures (from left to right: $BVI$, $BV$ and $VI$). \textit{Bottom}: BB radii for the same filter combinations.}
 \label{fig:BB}
\end{figure*}
\end{landscape}

The colour ($BV$, $BVI$ and $VI$) temperatures and BB radii versus rest-frame times relative to the date of explosion are presented in Fig.~\ref{fig:BB}.

\section{Distances}

\begin{table*}
\small
\centering
\setlength{\tabcolsep}{10pt}
\caption{Linear and Luminosity Distances to SE-SNe}
\label{table:distances}
\begin{tabular}{|cccccccc|}
\hline
SN	&	Type	&	Host	&	$z$	&		$\bar{D}$ (Mpc)$^{\dagger}$				&	$N_{\rm obs}$	&	$D_{\rm L}$ (Mpc)$^{\ddagger}$	&	Ref(s).	\\
\hline																			
2008ax	&	IIb	&	NGC 4490	&	0.0021	&	$	7.88	\pm	0.72	$	&	7	&	-	&	(1,13,19,53)	\\
2010as	&	IIb	&	NGC 6000	&	0.007354	&	$	26.13	\pm	2.54	$	&	6	&	-	&	(1,13,19)	\\
2011dh	&	IIb	&	M51	&	0.00155	&	$	7.63	\pm	0.07	$	&	30	&	-	&	(19-37)	\\
2011ei	&	IIb	&	NGC 6925	&	0.009317	&	$	30.07	\pm	30.13	$	&	18	&	-	&	(1,15-19,52)	\\
2011hs	&	IIb	&	IC 5267	&	0.005701	&	$	27.61	\pm	3.29	$	&	5	&	-	&	(15,19)	\\
1999dn	&	Ib	&	NGC 7714	&	0.00938	&	$	29.65	\pm	3.18	$	&	4	&	-	&	(1)	\\
2005bf	&	Ib	&	MCG+00-27-05	&	0.018913	&	$	88.10	\pm	9.45	$	&	4	&	-	&	(1)	\\
2008D	&	Ib	&	NGC 2770	&	0.0070	&	$	30.01	\pm	1.77	$	&	13	&	-	&	(1, 15-19)	\\
2009jf	&	Ib	&	NGC 7479	&	0.007942	&	$	33.58	\pm	3.34	$	&	4	&	-	&	(19)	\\
1994I	&	Ic	&	M51	&	0.00155	&	$	7.63	\pm	0.07	$	&	30	&	-	&	(19-37)	\\
2002ap	&	Ic	&	M74	&	0.002187	&	$	9.22	\pm	0.61	$	&	15	&	-	&	(19,26,44-51)	\\
2004aw	&	Ic	&	NGC 3997	&	0.0175	&	$	78.11	\pm	11.76	$	&	2	&	-	&	(1)	\\
2005ek	&	Ic	&	UGC 2526	&	0.016618	&	$	62.93	\pm	6.75	$	&	4	&	-	&	(65)	\\
2007gr	&	Ic	&	NGC 1058	&	0.001729	&	$	10.09	\pm	0.67	$	&	6	&	-	&	(16,19,38-42)	\\
2011bm	&	Ic	&	IC 3918	&	0.0221	&	$	129.38	\pm	5.16	$	&	6	&	-	&	(3,7,9,43)	\\
2009bb	&	IcBL	&	NGC 3278	&	0.009987	&	$	40.68	\pm	4.36	$	&	4	&	-	&	(1)	\\
2012ap	&	IcBL	&	NGC 1729	&	0.012241	&	$	40.37	\pm	7.38	$	&	2	&	-	&	(16)	\\
\hline																			
1998bw	&	GRB	&	ESO 184-G82	&	0.00867	&		-				&	-	&	37	&		\\
2003dh	&	GRB	&	ANON	&	0.1685	&		-				&	-	&	810	&		\\
2003lw	&	GRB	&	ANON	&	0.10536	&		-				&	-	&	487	&		\\
2006aj	&	GRB	&	ANON	&	0.03342	&		-				&	-	&	147	&		\\
2009nz	&	GRB	&	ANON	&	0.49	&		-				&	-	&	2765	&		\\
2010bh	&	GRB	&	ANON	&	0.0591	&		-				&	-	&	264	&		\\
2012bz	&	GRB	&	ANON	&	0.283	&		-				&	-	&	1452	&		\\
2013dx	&	GRB	&	ANON	&	0.145	&		-				&	-	&	687	&		\\
2016jca	&	GRB	&	ANON	&	0.1475	&		-				&	-	&	700	&		\\
\hline
\end{tabular}
\begin{flushleft}
$^{\dagger}$The weighted average linear distance and its associated error.\\
$^{\ddagger}$Luminosity distance calculated for a generic $\Lambda$CDM cosmology of $H_{0} = 70$ km s$^{-1}$ Mpc$^{-1}$, $\Omega_{\rm M} = 0.3$, $\Omega_{\Lambda} = 0.7$.\\ 
\scriptsize{\textbf{References}: (1) Theureau et al. (2007); (2) Wood-Vasey et al. (2008); (3) Mandel et al. (2011); (4) Mandel et al. (2009); (5)~Wang et al. (2006); (6) Takanashi et al. (2008); (7) Ganeshalingam et al. (2013); (8) Weyant et al. (2014); (9) Hicken et al. (2009); (10) Prieto et al. (2006); (11) Jha et al. (2007); (12) Reindl et al. (2005); (13) Terry et  al. (2002); (14) Sorce et al. (2012); (15) Willick et al. (1997); (16) Springob et al. (2009); (17) Tully et al. (2009); (18) Tully et al. (1992); (19) Tully (1988); (20) Ciardullo et al. (2002); (21) Ferrarese et al. (2000); (22) Feldmeier et al. (1997); (23) Tonry et al. (2001); (24)~Richmond et al. (1996); (25) Sofue (1991); (26) Zasov  \& Bizyaev (1996); (27) Bose  \& Kumar (2014); (28) Tak{\'a}ts  \& Vink{\'o} (2006); (29) Iwamoto et al. (1994); (30) Tak{\'a}ts  \& Vink{\'o} (2012); (31) Baron et al. (2007); (32) Baron et al. (1996); (33)~Poznanski et al. (2009); (34) Vink{\'o} et  al. (2012); (35) Dessart et al. (2008); (36) Tutui  \& Sofue (1997); (37) Chiba  \& Yoshii (1995); (38) Schmidt et al. (1994); (39) Schmidt et al. (1992); (40) Kirshner  \& Kwan (1974); (41) Zinn et  al. (2011); (42) Pierce (1994); (43) Amanullah et al. (2010); (44) Sohn  \& Davidge (1996); (45) Sharina et  al. (1996); (46) Hendry et al. (2005); (47) Herrmann et al.(2008); (48) Vink{\'o} et  al. (2004); (49) Van Dyk et al. (2006); (50) Olivares E.~et al. (2010); (51)~Jang  \& Lee (2014); (52) Pedreros  \& Madore (1981); (53) Karachentsev et al. (2013).}\\
\end{flushleft}
\end{table*}

\small
\onecolumn
\setlength{\tabcolsep}{4.0pt}
\begin{longtable}{|ccccccccc|} 
\caption{Empirically derived dilution factors ($\zeta$) of SE-SNe. NB: The values of the SE-SNe are derived from model-free distances to their host galaxies (see Table~\ref{table:distances}). Times are given in the rest-frame relative to the explosion epoch. The GRB-SN dilution factors have been calculated for luminosity distances derived from their spectroscopic redshifts for a generic, flat $\Lambda$CDM model ($H_{0} = 70$ km s$^{-1}$ Mpc$^{-1}$, $\Omega_{\rm M} = 0.3$, $\Omega_{\Lambda} = 0.7$).} \\
\hline
SN	&	Type	&	$t-t_0$ (d)$^{\dagger}$	&		$T_{BVI}$ (K)				&		$\zeta_{BVI}$				&		$T_{BV}$ (K)				&		$\zeta_{BV}$				&		$T_{VI}$ (K)				&		$\zeta_{VI}$	\\
\hline		
\endhead
1994I	&	Ic	&	7.63	&	$	31461	\pm	4052	$	&	$	0.275	\pm	0.054	$	&	$	22914	\pm	2951	$	&	$	0.355	\pm	0.070	$	&	$	34951	\pm	4501	$	&	$	0.256	\pm	0.050	$	\\
1994I	&	Ic	&	9.60	&	$	23138	\pm	2689	$	&	$	0.354	\pm	0.082	$	&	$	7504	\pm	872	$	&	$	1.379	\pm	0.320	$	&	$	23411	\pm	2720	$	&	$	0.351	\pm	0.081	$	\\
1994I	&	Ic	&	12.68	&	$	17916	\pm	6979	$	&	$	0.431	\pm	0.181	$	&	$	5731	\pm	2232	$	&	$	2.356	\pm	0.988	$	&	$	20945	\pm	8159	$	&	$	0.380	\pm	0.159	$	\\
1994I	&	Ic	&	14.60	&	$	8522	\pm	2175	$	&	$	0.856	\pm	0.411	$	&	$	5112	\pm	1305	$	&	$	2.792	\pm	1.340	$	&	$	14981	\pm	3824	$	&	$	0.470	\pm	0.226	$	\\
1994I	&	Ic	&	15.61	&	$	7431	\pm	2110	$	&	$	0.989	\pm	0.533	$	&	$	4114	\pm	1168	$	&	$	4.696	\pm	2.531	$	&	$	10937	\pm	3105	$	&	$	0.615	\pm	0.331	$	\\
1994I	&	Ic	&	15.73	&	$	7392	\pm	1623	$	&	$	0.998	\pm	0.474	$	&	$	4659	\pm	1022	$	&	$	3.290	\pm	1.563	$	&	$	11653	\pm	2558	$	&	$	0.576	\pm	0.274	$	\\
1994I	&	Ic	&	16.56	&	$	7574	\pm	2251	$	&	$	0.964	\pm	0.551	$	&	$	4114	\pm	1223	$	&	$	4.750	\pm	2.714	$	&	$	11464	\pm	3407	$	&	$	0.585	\pm	0.334	$	\\
1994I	&	Ic	&	17.54	&	$	7193	\pm	1988	$	&	$	1.012	\pm	0.607	$	&	$	4021	\pm	1112	$	&	$	4.882	\pm	2.928	$	&	$	10462	\pm	2892	$	&	$	0.629	\pm	0.378	$	\\
1994I	&	Ic	&	20.57	&	$	6762	\pm	1520	$	&	$	1.025	\pm	0.673	$	&	$	3846	\pm	864	$	&	$	4.860	\pm	3.192	$	&	$	8275	\pm	1860	$	&	$	0.768	\pm	0.505	$	\\
1994I	&	Ic	&	21.54	&	$	6359	\pm	1080	$	&	$	1.063	\pm	0.647	$	&	$	3931	\pm	668	$	&	$	4.243	\pm	2.583	$	&	$	7859	\pm	1335	$	&	$	0.788	\pm	0.480	$	\\
1994I	&	Ic	&	29.52	&	$	6758	\pm	187	$	&	$	0.878	\pm	0.579	$	&	$	4700	\pm	130	$	&	$	2.076	\pm	1.369	$	&	$	6792	\pm	188	$	&	$	0.872	\pm	0.575	$	\\
1994I	&	Ic	&	30.50	&	$	6734	\pm	35	$	&	$	0.836	\pm	0.566	$	&	$	5997	\pm	31	$	&	$	1.067	\pm	0.722	$	&	$	6736	\pm	35	$	&	$	0.836	\pm	0.566	$	\\
1994I	&	Ic	&	35.50	&	$	5969	\pm	58	$	&	$	1.152	\pm	0.925	$	&	$	6214	\pm	61	$	&	$	1.050	\pm	0.842	$	&	$	5917	\pm	58	$	&	$	1.169	\pm	0.938	$	\\
1999dn	&	Ib	&	7.64	&	$	-			$	&	$	-			$	&	$	5613	\pm	561	$	&	$	1.920	\pm	0.281	$	&	$	-			$	&	$	-			$	\\
1999dn	&	Ib	&	10.69	&	$	-			$	&	$	-			$	&	$	5123	\pm	512	$	&	$	1.903	\pm	0.279	$	&	$	-			$	&	$	-			$	\\
1999dn	&	Ib	&	16.61	&	$	-			$	&	$	-			$	&	$	4852	\pm	485	$	&	$	1.534	\pm	0.225	$	&	$	-			$	&	$	-			$	\\
1999dn	&	Ib	&	19.59	&	$	-			$	&	$	-			$	&	$	4533	\pm	453	$	&	$	1.820	\pm	0.267	$	&	$	-			$	&	$	-			$	\\
1999dn	&	Ib	&	25.25	&	$	-			$	&	$	-			$	&	$	3716	\pm	372	$	&	$	3.125	\pm	0.458	$	&	$	-			$	&	$	-			$	\\
1999dn	&	Ib	&	27.50	&	$	-			$	&	$	-			$	&	$	3622	\pm	362	$	&	$	2.964	\pm	0.435	$	&	$	-			$	&	$	-			$	\\
1999dn	&	Ib	&	28.24	&	$	-			$	&	$	-			$	&	$	3740	\pm	374	$	&	$	2.448	\pm	0.359	$	&	$	-			$	&	$	-			$	\\
1999dn	&	Ib	&	31.47	&	$	-			$	&	$	-			$	&	$	3470	\pm	347	$	&	$	2.790	\pm	0.409	$	&	$	-			$	&	$	-			$	\\
1999dn	&	Ib	&	50.99	&	$	-			$	&	$	-			$	&	$	3498	\pm	350	$	&	$	1.282	\pm	0.188	$	&	$	-			$	&	$	-			$	\\
2002ap	&	Ic	&	1.27	&	$	8016	\pm	988	$	&	$	0.537	\pm	0.117	$	&	$	5908	\pm	1210	$	&	$	1.081	\pm	0.250	$	&	$	10853	\pm	2370	$	&	$	0.378	\pm	0.082	$	\\
2002ap	&	Ic	&	2.23	&	$	11395	\pm	2277	$	&	$	0.425	\pm	0.120	$	&	$	7705	\pm	2088	$	&	$	0.844	\pm	0.248	$	&	$	20800	\pm	5879	$	&	$	0.249	\pm	0.070	$	\\
2002ap	&	Ic	&	3.21	&	$	11065	\pm	3246	$	&	$	0.507	\pm	0.214	$	&	$	7281	\pm	2998	$	&	$	1.080	\pm	0.468	$	&	$	27007	\pm	11419	$	&	$	0.232	\pm	0.098	$	\\
2002ap	&	Ic	&	4.34	&	$	10252	\pm	3898	$	&	$	0.599	\pm	0.328	$	&	$	6635	\pm	3568	$	&	$	1.322	\pm	0.737	$	&	$	21544	\pm	11797	$	&	$	0.285	\pm	0.156	$	\\
2002ap	&	Ic	&	5.21	&	$	10825	\pm	3232	$	&	$	0.533	\pm	0.222	$	&	$	6690	\pm	2698	$	&	$	1.316	\pm	0.562	$	&	$	24487	\pm	10168	$	&	$	0.259	\pm	0.108	$	\\
2002ap	&	Ic	&	6.34	&	$	9117	\pm	3091	$	&	$	0.717	\pm	0.387	$	&	$	6252	\pm	3323	$	&	$	1.506	\pm	0.825	$	&	$	17600	\pm	9501	$	&	$	0.345	\pm	0.186	$	\\
2002ap	&	Ic	&	7.20	&	$	9625	\pm	2002	$	&	$	0.583	\pm	0.189	$	&	$	6343	\pm	1990	$	&	$	1.375	\pm	0.460	$	&	$	18003	\pm	5838	$	&	$	0.318	\pm	0.103	$	\\
2002ap	&	Ic	&	8.21	&	$	9404	\pm	3326	$	&	$	0.643	\pm	0.348	$	&	$	5951	\pm	3186	$	&	$	1.631	\pm	0.893	$	&	$	19722	\pm	10681	$	&	$	0.302	\pm	0.164	$	\\
2002ap	&	Ic	&	9.69	&	$	8436	\pm	1647	$	&	$	0.760	\pm	0.247	$	&	$	5518	\pm	1678	$	&	$	2.028	\pm	0.699	$	&	$	14118	\pm	4588	$	&	$	0.440	\pm	0.143	$	\\
2002ap	&	Ic	&	10.33	&	$	8351	\pm	3116	$	&	$	0.787	\pm	0.493	$	&	$	5173	\pm	3203	$	&	$	2.330	\pm	1.478	$	&	$	17263	\pm	10822	$	&	$	0.350	\pm	0.220	$	\\
2002ap	&	Ic	&	11.21	&	$	7927	\pm	1767	$	&	$	0.849	\pm	0.370	$	&	$	5145	\pm	2113	$	&	$	2.516	\pm	1.158	$	&	$	16546	\pm	7217	$	&	$	0.389	\pm	0.170	$	\\
2002ap	&	Ic	&	12.20	&	$	7082	\pm	1701	$	&	$	1.088	\pm	0.567	$	&	$	4892	\pm	2423	$	&	$	2.933	\pm	1.601	$	&	$	13336	\pm	6952	$	&	$	0.500	\pm	0.260	$	\\
2002ap	&	Ic	&	13.20	&	$	7086	\pm	1551	$	&	$	1.073	\pm	0.528	$	&	$	4808	\pm	2201	$	&	$	3.104	\pm	1.629	$	&	$	13687	\pm	6741	$	&	$	0.490	\pm	0.241	$	\\
2002ap	&	Ic	&	14.20	&	$	7224	\pm	1275	$	&	$	1.085	\pm	0.386	$	&	$	4553	\pm	1307	$	&	$	3.881	\pm	1.605	$	&	$	12692	\pm	4517	$	&	$	0.573	\pm	0.204	$	\\
2002ap	&	Ic	&	15.20	&	$	6566	\pm	1375	$	&	$	1.292	\pm	0.661	$	&	$	4444	\pm	2039	$	&	$	4.075	\pm	2.278	$	&	$	11869	\pm	6069	$	&	$	0.608	\pm	0.311	$	\\
2002ap	&	Ic	&	16.33	&	$	6400	\pm	1993	$	&	$	1.624	\pm	1.163	$	&	$	4226	\pm	2798	$	&	$	5.288	\pm	4.053	$	&	$	11299	\pm	8093	$	&	$	0.721	\pm	0.516	$	\\
2002ap	&	Ic	&	17.19	&	$	6274	\pm	828	$	&	$	1.396	\pm	0.520	$	&	$	4261	\pm	1102	$	&	$	4.593	\pm	2.110	$	&	$	9765	\pm	3640	$	&	$	0.783	\pm	0.292	$	\\
2002ap	&	Ic	&	18.19	&	$	5989	\pm	1142	$	&	$	1.502	\pm	0.794	$	&	$	4208	\pm	1852	$	&	$	4.511	\pm	2.727	$	&	$	9303	\pm	4920	$	&	$	0.795	\pm	0.420	$	\\
2002ap	&	Ic	&	19.31	&	$	6008	\pm	1719	$	&	$	2.339	\pm	2.127	$	&	$	4000	\pm	2564	$	&	$	7.983	\pm	8.901	$	&	$	9804	\pm	8916	$	&	$	1.107	\pm	1.007	$	\\
2002ap	&	Ic	&	20.46	&	$	6339	\pm	1629	$	&	$	1.726	\pm	1.447	$	&	$	3907	\pm	1949	$	&	$	7.382	\pm	7.945	$	&	$	9676	\pm	8116	$	&	$	0.965	\pm	0.810	$	\\
2002ap	&	Ic	&	23.19	&	$	5525	\pm	922	$	&	$	2.148	\pm	1.973	$	&	$	3903	\pm	1626	$	&	$	6.994	\pm	8.607	$	&	$	8400	\pm	7717	$	&	$	1.133	\pm	1.040	$	\\
2002ap	&	Ic	&	24.18	&	$	5454	\pm	980	$	&	$	2.096	\pm	1.946	$	&	$	3970	\pm	1818	$	&	$	6.060	\pm	7.456	$	&	$	8002	\pm	7429	$	&	$	1.126	\pm	1.045	$	\\
2002ap	&	Ic	&	25.61	&	$	5338	\pm	666	$	&	$	2.557	\pm	2.835	$	&	$	3776	\pm	1198	$	&	$	8.768	\pm	13.463	$	&	$	7962	\pm	8827	$	&	$	1.387	\pm	1.538	$	\\
2004aw	&	Ic	&	11.15	&	$	7494	\pm	1356	$	&	$	1.549	\pm	0.618	$	&	$	5297	\pm	959	$	&	$	3.559	\pm	1.421	$	&	$	9670	\pm	1750	$	&	$	1.106	\pm	0.442	$	\\
2004aw	&	Ic	&	12.14	&	$	6843	\pm	840	$	&	$	1.760	\pm	0.623	$	&	$	5381	\pm	660	$	&	$	3.286	\pm	1.163	$	&	$	8689	\pm	1066	$	&	$	1.253	\pm	0.443	$	\\
2004aw	&	Ic	&	12.83	&	$	7024	\pm	730	$	&	$	1.648	\pm	0.522	$	&	$	5652	\pm	588	$	&	$	2.748	\pm	0.870	$	&	$	7909	\pm	822	$	&	$	1.383	\pm	0.438	$	\\
2004aw	&	Ic	&	13.81	&	$	6453	\pm	759	$	&	$	1.912	\pm	0.650	$	&	$	4985	\pm	586	$	&	$	3.787	\pm	1.286	$	&	$	7539	\pm	887	$	&	$	1.511	\pm	0.513	$	\\
2004aw	&	Ic	&	14.79	&	$	5888	\pm	710	$	&	$	2.221	\pm	0.861	$	&	$	4583	\pm	553	$	&	$	4.810	\pm	1.864	$	&	$	7756	\pm	936	$	&	$	1.440	\pm	0.558	$	\\
2004aw	&	Ic	&	15.68	&	$	5669	\pm	678	$	&	$	2.231	\pm	0.773	$	&	$	4018	\pm	480	$	&	$	7.032	\pm	2.437	$	&	$	7756	\pm	927	$	&	$	1.397	\pm	0.484	$	\\
2004aw	&	Ic	&	21.47	&	$	5213	\pm	597	$	&	$	2.153	\pm	0.856	$	&	$	4114	\pm	471	$	&	$	4.817	\pm	1.915	$	&	$	6526	\pm	747	$	&	$	1.442	\pm	0.573	$	\\
2004aw	&	Ic	&	21.67	&	$	4932	\pm	495	$	&	$	2.402	\pm	0.937	$	&	$	3903	\pm	392	$	&	$	5.664	\pm	2.210	$	&	$	6282	\pm	630	$	&	$	1.542	\pm	0.602	$	\\
2004aw	&	Ic	&	22.95	&	$	5486	\pm	705	$	&	$	1.838	\pm	0.715	$	&	$	4114	\pm	529	$	&	$	4.442	\pm	1.728	$	&	$	6282	\pm	808	$	&	$	1.441	\pm	0.561	$	\\
2004aw	&	Ic	&	26.39	&	$	5090	\pm	625	$	&	$	1.915	\pm	0.764	$	&	$	3903	\pm	479	$	&	$	4.616	\pm	1.843	$	&	$	5891	\pm	723	$	&	$	1.450	\pm	0.579	$	\\
2004aw	&	Ic	&	28.45	&	$	4634	\pm	532	$	&	$	2.246	\pm	1.006	$	&	$	3881	\pm	446	$	&	$	4.325	\pm	1.938	$	&	$	5733	\pm	659	$	&	$	1.419	\pm	0.636	$	\\
2004aw	&	Ic	&	29.83	&	$	5235	\pm	564	$	&	$	1.552	\pm	0.547	$	&	$	3903	\pm	420	$	&	$	3.988	\pm	1.404	$	&	$	5851	\pm	630	$	&	$	1.272	\pm	0.448	$	\\
2004aw	&	Ic	&	31.40	&	$	5088	\pm	570	$	&	$	1.561	\pm	0.577	$	&	$	3838	\pm	430	$	&	$	3.982	\pm	1.471	$	&	$	5772	\pm	647	$	&	$	1.237	\pm	0.457	$	\\
2004aw	&	Ic	&	34.35	&	$	4637	\pm	445	$	&	$	1.722	\pm	0.660	$	&	$	3795	\pm	364	$	&	$	3.592	\pm	1.376	$	&	$	5411	\pm	519	$	&	$	1.254	\pm	0.480	$	\\
2005bf	&	Ib	&	8.00	&	$	-			$	&	$	-			$	&	$	8817	\pm	882	$	&	$	1.147	\pm	0.168	$	&	$	-			$	&	$	-			$	\\
2005bf	&	Ib	&	8.91	&	$	-			$	&	$	-			$	&	$	9442	\pm	944	$	&	$	0.990	\pm	0.145	$	&	$	-			$	&	$	-			$	\\
2005bf	&	Ib	&	12.85	&	$	-			$	&	$	-			$	&	$	9880	\pm	988	$	&	$	0.777	\pm	0.114	$	&	$	-			$	&	$	-			$	\\
2005bf	&	Ib	&	13.85	&	$	-			$	&	$	-			$	&	$	9653	\pm	965	$	&	$	0.759	\pm	0.111	$	&	$	-			$	&	$	-			$	\\
2005bf	&	Ib	&	14.79	&	$	-			$	&	$	-			$	&	$	8978	\pm	898	$	&	$	0.799	\pm	0.117	$	&	$	-			$	&	$	-			$	\\
2005bf	&	Ib	&	15.84	&	$	-			$	&	$	-			$	&	$	8813	\pm	881	$	&	$	0.761	\pm	0.112	$	&	$	-			$	&	$	-			$	\\
2005bf	&	Ib	&	16.74	&	$	-			$	&	$	-			$	&	$	8317	\pm	832	$	&	$	0.779	\pm	0.114	$	&	$	-			$	&	$	-			$	\\
2005bf	&	Ib	&	17.72	&	$	-			$	&	$	-			$	&	$	7904	\pm	790	$	&	$	0.794	\pm	0.116	$	&	$	-			$	&	$	-			$	\\
2005bf	&	Ib	&	18.72	&	$	-			$	&	$	-			$	&	$	7239	\pm	724	$	&	$	0.887	\pm	0.130	$	&	$	-			$	&	$	-			$	\\
2005bf	&	Ib	&	19.65	&	$	-			$	&	$	-			$	&	$	6848	\pm	685	$	&	$	0.965	\pm	0.142	$	&	$	-			$	&	$	-			$	\\
2005bf	&	Ib	&	20.65	&	$	-			$	&	$	-			$	&	$	6641	\pm	664	$	&	$	0.990	\pm	0.145	$	&	$	-			$	&	$	-			$	\\
2005bf	&	Ib	&	21.60	&	$	-			$	&	$	-			$	&	$	6513	\pm	651	$	&	$	0.997	\pm	0.146	$	&	$	-			$	&	$	-			$	\\
2005bf	&	Ib	&	22.59	&	$	-			$	&	$	-			$	&	$	6485	\pm	649	$	&	$	0.991	\pm	0.145	$	&	$	-			$	&	$	-			$	\\
2005bf	&	Ib	&	23.59	&	$	-			$	&	$	-			$	&	$	6608	\pm	661	$	&	$	0.936	\pm	0.137	$	&	$	-			$	&	$	-			$	\\
2005bf	&	Ib	&	24.64	&	$	-			$	&	$	-			$	&	$	6411	\pm	641	$	&	$	0.979	\pm	0.143	$	&	$	-			$	&	$	-			$	\\
2005bf	&	Ib	&	29.45	&	$	-			$	&	$	-			$	&	$	7339	\pm	734	$	&	$	0.742	\pm	0.109	$	&	$	-			$	&	$	-			$	\\
2005bf	&	Ib	&	30.42	&	$	-			$	&	$	-			$	&	$	7687	\pm	769	$	&	$	0.688	\pm	0.101	$	&	$	-			$	&	$	-			$	\\
2005bf	&	Ib	&	31.39	&	$	-			$	&	$	-			$	&	$	7881	\pm	788	$	&	$	0.665	\pm	0.097	$	&	$	-			$	&	$	-			$	\\
2005bf	&	Ib	&	48.24	&	$	-			$	&	$	-			$	&	$	5751	\pm	575	$	&	$	1.031	\pm	0.151	$	&	$	-			$	&	$	-			$	\\
2005bf	&	Ib	&	49.03	&	$	-			$	&	$	-			$	&	$	5495	\pm	549	$	&	$	1.129	\pm	0.165	$	&	$	-			$	&	$	-			$	\\
2005bf	&	Ib	&	50.05	&	$	-			$	&	$	-			$	&	$	5357	\pm	536	$	&	$	1.169	\pm	0.172	$	&	$	-			$	&	$	-			$	\\
2005bf	&	Ib	&	51.02	&	$	-			$	&	$	-			$	&	$	5134	\pm	513	$	&	$	1.256	\pm	0.184	$	&	$	-			$	&	$	-			$	\\
2005bf	&	Ib	&	52.16	&	$	-			$	&	$	-			$	&	$	4818	\pm	482	$	&	$	1.429	\pm	0.209	$	&	$	-			$	&	$	-			$	\\
2005bf	&	Ib	&	52.96	&	$	-			$	&	$	-			$	&	$	4966	\pm	497	$	&	$	1.280	\pm	0.188	$	&	$	-			$	&	$	-			$	\\
2005bf	&	Ib	&	61.78	&	$	-			$	&	$	-			$	&	$	4071	\pm	407	$	&	$	1.540	\pm	0.226	$	&	$	-			$	&	$	-			$	\\
2005bf	&	Ib	&	62.78	&	$	-			$	&	$	-			$	&	$	3985	\pm	399	$	&	$	1.553	\pm	0.228	$	&	$	-			$	&	$	-			$	\\
2005bf	&	Ib	&	63.74	&	$	-			$	&	$	-			$	&	$	3839	\pm	384	$	&	$	1.684	\pm	0.247	$	&	$	-			$	&	$	-			$	\\
2005ek	&	Ic	&	5.12	&	$	8104	\pm	53	$	&	$	2.336	\pm	0.319	$	&	$	7978	\pm	52	$	&	$	2.406	\pm	0.329	$	&	$	8174	\pm	53	$	&	$	2.309	\pm	0.316	$	\\
2005ek	&	Ic	&	6.10	&	$	7237	\pm	758	$	&	$	2.238	\pm	0.569	$	&	$	8994	\pm	943	$	&	$	1.487	\pm	0.378	$	&	$	6436	\pm	674	$	&	$	2.727	\pm	0.694	$	\\
2005ek	&	Ic	&	7.08	&	$	6570	\pm	750	$	&	$	2.345	\pm	0.675	$	&	$	5399	\pm	617	$	&	$	3.756	\pm	1.082	$	&	$	7351	\pm	840	$	&	$	1.934	\pm	0.557	$	\\
2005ek	&	Ic	&	8.95	&	$	6420	\pm	855	$	&	$	1.936	\pm	0.619	$	&	$	4770	\pm	635	$	&	$	4.132	\pm	1.322	$	&	$	7158	\pm	953	$	&	$	1.613	\pm	0.516	$	\\
2005ek	&	Ic	&	11.02	&	$	5808	\pm	613	$	&	$	1.927	\pm	0.567	$	&	$	4080	\pm	431	$	&	$	5.271	\pm	1.551	$	&	$	6199	\pm	655	$	&	$	1.719	\pm	0.506	$	\\
2005ek	&	Ic	&	12.20	&	$	5362	\pm	483	$	&	$	1.974	\pm	0.610	$	&	$	3820	\pm	344	$	&	$	5.574	\pm	1.723	$	&	$	5670	\pm	511	$	&	$	1.782	\pm	0.551	$	\\
2005ek	&	Ic	&	14.85	&	$	4573	\pm	158	$	&	$	2.016	\pm	0.699	$	&	$	4111	\pm	142	$	&	$	2.884	\pm	1.000	$	&	$	4766	\pm	164	$	&	$	1.847	\pm	0.641	$	\\
2005ek	&	Ic	&	16.03	&	$	4607	\pm	171	$	&	$	1.705	\pm	0.655	$	&	$	4111	\pm	152	$	&	$	2.501	\pm	0.960	$	&	$	4818	\pm	179	$	&	$	1.554	\pm	0.596	$	\\
2007gr	&	Ic	&	5.06	&	$	9932	\pm	561	$	&	$	1.395	\pm	0.162	$	&	$	8909	\pm	503	$	&	$	1.677	\pm	0.195	$	&	$	10938	\pm	618	$	&	$	1.249	\pm	0.145	$	\\
2007gr	&	Ic	&	7.19	&	$	8922	\pm	1623	$	&	$	1.511	\pm	0.482	$	&	$	7369	\pm	1340	$	&	$	2.189	\pm	0.699	$	&	$	13851	\pm	2520	$	&	$	0.889	\pm	0.284	$	\\
2007gr	&	Ic	&	8.22	&	$	9280	\pm	2324	$	&	$	1.323	\pm	0.555	$	&	$	7171	\pm	1796	$	&	$	2.182	\pm	0.916	$	&	$	19083	\pm	4780	$	&	$	0.609	\pm	0.256	$	\\
2007gr	&	Ic	&	9.11	&	$	9550	\pm	1324	$	&	$	1.182	\pm	0.259	$	&	$	7404	\pm	1026	$	&	$	1.903	\pm	0.418	$	&	$	12418	\pm	1721	$	&	$	0.885	\pm	0.194	$	\\
2007gr	&	Ic	&	10.22	&	$	8452	\pm	1870	$	&	$	1.332	\pm	0.519	$	&	$	6667	\pm	1475	$	&	$	2.159	\pm	0.841	$	&	$	13546	\pm	2998	$	&	$	0.742	\pm	0.289	$	\\
2007gr	&	Ic	&	11.08	&	$	8650	\pm	955	$	&	$	1.238	\pm	0.243	$	&	$	7053	\pm	779	$	&	$	1.863	\pm	0.365	$	&	$	10601	\pm	1171	$	&	$	0.968	\pm	0.190	$	\\
2007gr	&	Ic	&	13.01	&	$	8686	\pm	1654	$	&	$	1.111	\pm	0.346	$	&	$	6151	\pm	1171	$	&	$	2.326	\pm	0.724	$	&	$	12954	\pm	2467	$	&	$	0.711	\pm	0.221	$	\\
2007gr	&	Ic	&	14.08	&	$	7939	\pm	1177	$	&	$	1.309	\pm	0.354	$	&	$	6094	\pm	903	$	&	$	2.364	\pm	0.639	$	&	$	10571	\pm	1567	$	&	$	0.915	\pm	0.247	$	\\
2007gr	&	Ic	&	16.84	&	$	6916	\pm	745	$	&	$	1.627	\pm	0.396	$	&	$	5783	\pm	623	$	&	$	2.544	\pm	0.619	$	&	$	8720	\pm	939	$	&	$	1.161	\pm	0.282	$	\\
2007gr	&	Ic	&	18.75	&	$	6588	\pm	701	$	&	$	1.651	\pm	0.384	$	&	$	5340	\pm	568	$	&	$	2.856	\pm	0.664	$	&	$	8041	\pm	855	$	&	$	1.231	\pm	0.286	$	\\
2007gr	&	Ic	&	21.73	&	$	6121	\pm	786	$	&	$	1.607	\pm	0.471	$	&	$	4738	\pm	609	$	&	$	3.363	\pm	0.985	$	&	$	7911	\pm	1016	$	&	$	1.088	\pm	0.319	$	\\
2007gr	&	Ic	&	24.95	&	$	5351	\pm	424	$	&	$	1.771	\pm	0.378	$	&	$	4527	\pm	359	$	&	$	2.993	\pm	0.639	$	&	$	6194	\pm	491	$	&	$	1.363	\pm	0.291	$	\\
2007gr	&	Ic	&	28.12	&	$	4898	\pm	371	$	&	$	1.868	\pm	0.434	$	&	$	4274	\pm	324	$	&	$	2.965	\pm	0.689	$	&	$	5630	\pm	427	$	&	$	1.410	\pm	0.328	$	\\
2008ax	&	IIB	&	4.79	&	$	8934	\pm	1365	$	&	$	0.717	\pm	0.177	$	&	$	6000	\pm	917	$	&	$	1.616	\pm	0.399	$	&	$	11426	\pm	1746	$	&	$	0.547	\pm	0.135	$	\\
2008ax	&	IIB	&	8.68	&	$	12838	\pm	2406	$	&	$	0.369	\pm	0.088	$	&	$	8135	\pm	1524	$	&	$	0.739	\pm	0.176	$	&	$	17323	\pm	3246	$	&	$	0.284	\pm	0.068	$	\\
2008ax	&	IIB	&	10.88	&	$	14241	\pm	3228	$	&	$	0.343	\pm	0.089	$	&	$	8416	\pm	1908	$	&	$	0.732	\pm	0.190	$	&	$	20825	\pm	4720	$	&	$	0.251	\pm	0.065	$	\\
2008ax	&	IIB	&	13.57	&	$	14387	\pm	1880	$	&	$	0.315	\pm	0.059	$	&	$	10229	\pm	1337	$	&	$	0.494	\pm	0.092	$	&	$	17323	\pm	2264	$	&	$	0.269	\pm	0.050	$	\\
2008ax	&	IIB	&	27.54	&	$	7459	\pm	1012	$	&	$	0.651	\pm	0.164	$	&	$	5164	\pm	701	$	&	$	1.564	\pm	0.393	$	&	$	9254	\pm	1256	$	&	$	0.496	\pm	0.125	$	\\
2008ax	&	IIB	&	30.64	&	$	6345	\pm	909	$	&	$	0.753	\pm	0.219	$	&	$	4280	\pm	613	$	&	$	2.300	\pm	0.668	$	&	$	8082	\pm	1158	$	&	$	0.536	\pm	0.156	$	\\
2008ax	&	IIB	&	31.73	&	$	6008	\pm	834	$	&	$	0.821	\pm	0.242	$	&	$	4082	\pm	567	$	&	$	2.604	\pm	0.767	$	&	$	7612	\pm	1057	$	&	$	0.580	\pm	0.171	$	\\
2008ax	&	IIB	&	32.73	&	$	5810	\pm	796	$	&	$	0.850	\pm	0.254	$	&	$	3960	\pm	542	$	&	$	2.770	\pm	0.827	$	&	$	7331	\pm	1004	$	&	$	0.598	\pm	0.179	$	\\
2008ax	&	IIB	&	40.51	&	$	5277	\pm	601	$	&	$	0.869	\pm	0.241	$	&	$	3790	\pm	431	$	&	$	2.586	\pm	0.717	$	&	$	6356	\pm	724	$	&	$	0.637	\pm	0.176	$	\\
2008ax	&	IIB	&	42.81	&	$	5123	\pm	551	$	&	$	0.854	\pm	0.231	$	&	$	3967	\pm	427	$	&	$	0.786	\pm	0.212	$	&	$	6166	\pm	663	$	&	$	0.622	\pm	0.168	$	\\
2008ax	&	IIB	&	53.49	&	$	5319	\pm	535	$	&	$	0.626	\pm	0.158	$	&	$	3960	\pm	398	$	&	$	1.609	\pm	0.405	$	&	$	6212	\pm	625	$	&	$	0.482	\pm	0.121	$	\\
2008D	&	Ib	&	3.43	&	$	8378	\pm	35	$	&	$	0.814	\pm	0.092	$	&	$	8504	\pm	36	$	&	$	0.793	\pm	0.090	$	&	$	8337	\pm	35	$	&	$	0.819	\pm	0.093	$	\\
2008D	&	Ib	&	5.63	&	$	8140	\pm	1426	$	&	$	0.597	\pm	0.166	$	&	$	5189	\pm	909	$	&	$	1.677	\pm	0.467	$	&	$	11039	\pm	1933	$	&	$	0.421	\pm	0.117	$	\\
2008D	&	Ib	&	6.43	&	$	8107	\pm	738	$	&	$	0.563	\pm	0.098	$	&	$	6264	\pm	570	$	&	$	0.961	\pm	0.168	$	&	$	9221	\pm	839	$	&	$	0.481	\pm	0.084	$	\\
2008D	&	Ib	&	7.64	&	$	8048	\pm	826	$	&	$	0.530	\pm	0.100	$	&	$	6041	\pm	620	$	&	$	0.972	\pm	0.183	$	&	$	9337	\pm	958	$	&	$	0.442	\pm	0.083	$	\\
2008D	&	Ib	&	8.39	&	$	7403	\pm	946	$	&	$	0.616	\pm	0.142	$	&	$	5206	\pm	665	$	&	$	1.425	\pm	0.328	$	&	$	9084	\pm	1161	$	&	$	0.475	\pm	0.109	$	\\
2008D	&	Ib	&	16.22	&	$	8172	\pm	1254	$	&	$	0.512	\pm	0.123	$	&	$	5329	\pm	818	$	&	$	1.343	\pm	0.323	$	&	$	10652	\pm	1634	$	&	$	0.378	\pm	0.091	$	\\
2008D	&	Ib	&	19.35	&	$	6999	\pm	511	$	&	$	0.611	\pm	0.101	$	&	$	5646	\pm	412	$	&	$	1.008	\pm	0.167	$	&	$	7739	\pm	565	$	&	$	0.532	\pm	0.088	$	\\
2008D	&	Ib	&	21.27	&	$	6825	\pm	494	$	&	$	0.623	\pm	0.104	$	&	$	5508	\pm	399	$	&	$	1.038	\pm	0.173	$	&	$	7539	\pm	546	$	&	$	0.541	\pm	0.090	$	\\
2008D	&	Ib	&	22.36	&	$	6509	\pm	498	$	&	$	0.654	\pm	0.115	$	&	$	5194	\pm	397	$	&	$	1.154	\pm	0.203	$	&	$	7245	\pm	554	$	&	$	0.560	\pm	0.099	$	\\
2008D	&	Ib	&	23.28	&	$	6308	\pm	499	$	&	$	0.670	\pm	0.122	$	&	$	4996	\pm	395	$	&	$	1.231	\pm	0.225	$	&	$	7054	\pm	558	$	&	$	0.568	\pm	0.104	$	\\
2008D	&	Ib	&	25.30	&	$	6058	\pm	470	$	&	$	0.671	\pm	0.124	$	&	$	4812	\pm	373	$	&	$	1.251	\pm	0.231	$	&	$	6759	\pm	524	$	&	$	0.566	\pm	0.105	$	\\
2008D	&	Ib	&	32.25	&	$	4983	\pm	476	$	&	$	0.697	\pm	0.169	$	&	$	3752	\pm	358	$	&	$	1.826	\pm	0.442	$	&	$	5770	\pm	551	$	&	$	0.536	\pm	0.130	$	\\
2008D	&	Ib	&	40.13	&	$	4478	\pm	381	$	&	$	0.623	\pm	0.149	$	&	$	3457	\pm	294	$	&	$	1.629	\pm	0.390	$	&	$	5083	\pm	433	$	&	$	0.484	\pm	0.116	$	\\
2009bb	&	IcBL	&	10.89	&	$	10939	\pm	2154	$	&	$	0.464	\pm	0.121	$	&	$	7158	\pm	1410	$	&	$	0.997	\pm	0.260	$	&	$	16933	\pm	3335	$	&	$	0.309	\pm	0.081	$	\\
2009bb	&	IcBL	&	17.82	&	$	7719	\pm	1671	$	&	$	0.643	\pm	0.221	$	&	$	4818	\pm	1043	$	&	$	2.102	\pm	0.722	$	&	$	11614	\pm	2514	$	&	$	0.401	\pm	0.138	$	\\
2009bb	&	IcBL	&	18.72	&	$	6767	\pm	1237	$	&	$	0.762	\pm	0.259	$	&	$	4467	\pm	816	$	&	$	2.468	\pm	0.837	$	&	$	10080	\pm	1842	$	&	$	0.458	\pm	0.156	$	\\
2009bb	&	IcBL	&	20.70	&	$	6179	\pm	872	$	&	$	0.849	\pm	0.247	$	&	$	4380	\pm	618	$	&	$	2.372	\pm	0.690	$	&	$	8321	\pm	1174	$	&	$	0.557	\pm	0.162	$	\\
2009bb	&	IcBL	&	21.69	&	$	6017	\pm	844	$	&	$	0.865	\pm	0.256	$	&	$	4304	\pm	604	$	&	$	2.389	\pm	0.707	$	&	$	7903	\pm	1109	$	&	$	0.580	\pm	0.172	$	\\
2009bb	&	IcBL	&	23.67	&	$	5651	\pm	781	$	&	$	0.908	\pm	0.278	$	&	$	4054	\pm	560	$	&	$	2.643	\pm	0.809	$	&	$	7348	\pm	1015	$	&	$	0.605	\pm	0.185	$	\\
2009bb	&	IcBL	&	27.73	&	$	5339	\pm	623	$	&	$	0.851	\pm	0.237	$	&	$	3970	\pm	463	$	&	$	2.300	\pm	0.639	$	&	$	6720	\pm	784	$	&	$	0.584	\pm	0.162	$	\\
2009bb	&	IcBL	&	28.72	&	$	5310	\pm	668	$	&	$	0.821	\pm	0.242	$	&	$	3879	\pm	488	$	&	$	2.382	\pm	0.703	$	&	$	6758	\pm	850	$	&	$	0.553	\pm	0.163	$	\\
2009bb	&	IcBL	&	32.68	&	$	5342	\pm	618	$	&	$	0.688	\pm	0.190	$	&	$	3993	\pm	462	$	&	$	1.814	\pm	0.500	$	&	$	6649	\pm	769	$	&	$	0.480	\pm	0.132	$	\\
2009bb	&	IcBL	&	37.53	&	$	5451	\pm	651	$	&	$	0.575	\pm	0.159	$	&	$	4052	\pm	484	$	&	$	1.506	\pm	0.418	$	&	$	6721	\pm	803	$	&	$	0.409	\pm	0.113	$	\\
2009jf	&	Ib	&	13.78	&	$	8615	\pm	1210	$	&	$	0.745	\pm	0.194	$	&	$	6572	\pm	923	$	&	$	1.265	\pm	0.330	$	&	$	10269	\pm	1442	$	&	$	0.599	\pm	0.156	$	\\
2009jf	&	Ib	&	15.76	&	$	8415	\pm	1047	$	&	$	0.725	\pm	0.179	$	&	$	6546	\pm	814	$	&	$	1.207	\pm	0.298	$	&	$	10255	\pm	1276	$	&	$	0.568	\pm	0.140	$	\\
2009jf	&	Ib	&	16.92	&	$	8777	\pm	875	$	&	$	0.665	\pm	0.134	$	&	$	6853	\pm	684	$	&	$	1.066	\pm	0.215	$	&	$	9968	\pm	994	$	&	$	0.572	\pm	0.115	$	\\
2009jf	&	Ib	&	17.76	&	$	8834	\pm	846	$	&	$	0.655	\pm	0.128	$	&	$	6760	\pm	647	$	&	$	1.073	\pm	0.209	$	&	$	9557	\pm	915	$	&	$	0.596	\pm	0.116	$	\\
2009jf	&	Ib	&	19.68	&	$	7967	\pm	903	$	&	$	0.735	\pm	0.173	$	&	$	6201	\pm	703	$	&	$	1.251	\pm	0.294	$	&	$	9437	\pm	1069	$	&	$	0.592	\pm	0.139	$	\\
2009jf	&	Ib	&	19.94	&	$	8900	\pm	1134	$	&	$	0.638	\pm	0.140	$	&	$	5810	\pm	740	$	&	$	1.461	\pm	0.321	$	&	$	9662	\pm	1231	$	&	$	0.578	\pm	0.127	$	\\
2009jf	&	Ib	&	22.85	&	$	7958	\pm	975	$	&	$	0.716	\pm	0.172	$	&	$	5827	\pm	714	$	&	$	1.356	\pm	0.326	$	&	$	8673	\pm	1062	$	&	$	0.637	\pm	0.153	$	\\
2009jf	&	Ib	&	25.81	&	$	6823	\pm	714	$	&	$	0.816	\pm	0.207	$	&	$	5026	\pm	526	$	&	$	1.822	\pm	0.464	$	&	$	8948	\pm	937	$	&	$	0.568	\pm	0.145	$	\\
2009jf	&	Ib	&	28.61	&	$	6364	\pm	782	$	&	$	0.866	\pm	0.253	$	&	$	4809	\pm	591	$	&	$	1.852	\pm	0.542	$	&	$	7898	\pm	971	$	&	$	0.629	\pm	0.184	$	\\
2009jf	&	Ib	&	31.79	&	$	6279	\pm	944	$	&	$	0.812	\pm	0.260	$	&	$	4357	\pm	655	$	&	$	2.215	\pm	0.709	$	&	$	7487	\pm	1126	$	&	$	0.622	\pm	0.199	$	\\
2009jf	&	Ib	&	39.52	&	$	4900	\pm	614	$	&	$	0.957	\pm	0.338	$	&	$	3649	\pm	457	$	&	$	2.683	\pm	0.948	$	&	$	5953	\pm	746	$	&	$	0.666	\pm	0.235	$	\\
2009jf	&	Ib	&	40.56	&	$	4706	\pm	538	$	&	$	0.991	\pm	0.356	$	&	$	3584	\pm	410	$	&	$	2.707	\pm	0.973	$	&	$	5882	\pm	672	$	&	$	0.647	\pm	0.233	$	\\
2009jf	&	Ib	&	47.45	&	$	4804	\pm	678	$	&	$	0.717	\pm	0.286	$	&	$	3581	\pm	505	$	&	$	1.981	\pm	0.791	$	&	$	5668	\pm	800	$	&	$	0.513	\pm	0.205	$	\\
2009jf	&	Ib	&	49.45	&	$	4866	\pm	623	$	&	$	0.637	\pm	0.223	$	&	$	3582	\pm	458	$	&	$	1.846	\pm	0.648	$	&	$	5748	\pm	736	$	&	$	0.464	\pm	0.163	$	\\
2009jf	&	Ib	&	54.61	&	$	4589	\pm	420	$	&	$	0.619	\pm	0.183	$	&	$	3737	\pm	342	$	&	$	1.289	\pm	0.381	$	&	$	5252	\pm	480	$	&	$	0.468	\pm	0.138	$	\\
2009jf	&	Ib	&	59.38	&	$	4749	\pm	337	$	&	$	0.483	\pm	0.098	$	&	$	3614	\pm	256	$	&	$	1.296	\pm	0.262	$	&	$	5563	\pm	394	$	&	$	0.363	\pm	0.073	$	\\
2010as	&	IIB	&	4.42	&	$	8929	\pm	2168	$	&	$	1.473	\pm	0.466	$	&	$	4626	\pm	1123	$	&	$	6.962	\pm	2.204	$	&	$	16345	\pm	3969	$	&	$	0.820	\pm	0.260	$	\\
2010as	&	IIB	&	5.41	&	$	9596	\pm	1461	$	&	$	1.255	\pm	0.275	$	&	$	6064	\pm	924	$	&	$	3.101	\pm	0.680	$	&	$	12670	\pm	1930	$	&	$	0.947	\pm	0.208	$	\\
2010as	&	IIB	&	6.40	&	$	9095	\pm	1210	$	&	$	1.269	\pm	0.263	$	&	$	6007	\pm	799	$	&	$	2.938	\pm	0.608	$	&	$	11500	\pm	1530	$	&	$	0.989	\pm	0.205	$	\\
2010as	&	IIB	&	8.29	&	$	10219	\pm	1952	$	&	$	0.986	\pm	0.243	$	&	$	5924	\pm	1132	$	&	$	2.851	\pm	0.703	$	&	$	15020	\pm	2870	$	&	$	0.684	\pm	0.169	$	\\
2010as	&	IIB	&	9.38	&	$	10468	\pm	1384	$	&	$	0.911	\pm	0.179	$	&	$	6951	\pm	919	$	&	$	1.885	\pm	0.370	$	&	$	13132	\pm	1736	$	&	$	0.730	\pm	0.143	$	\\
2010as	&	IIB	&	10.47	&	$	11145	\pm	1936	$	&	$	0.779	\pm	0.174	$	&	$	6728	\pm	1169	$	&	$	1.895	\pm	0.424	$	&	$	15446	\pm	2683	$	&	$	0.577	\pm	0.129	$	\\
2010as	&	IIB	&	12.36	&	$	11057	\pm	1746	$	&	$	0.755	\pm	0.160	$	&	$	6912	\pm	1092	$	&	$	1.712	\pm	0.363	$	&	$	14718	\pm	2324	$	&	$	0.579	\pm	0.123	$	\\
2010as	&	IIB	&	14.34	&	$	10668	\pm	1597	$	&	$	0.767	\pm	0.160	$	&	$	6791	\pm	1017	$	&	$	1.716	\pm	0.357	$	&	$	13946	\pm	2088	$	&	$	0.594	\pm	0.124	$	\\
2010as	&	IIB	&	15.34	&	$	10362	\pm	1479	$	&	$	0.792	\pm	0.162	$	&	$	6702	\pm	956	$	&	$	1.747	\pm	0.358	$	&	$	13339	\pm	1904	$	&	$	0.619	\pm	0.127	$	\\
2010as	&	IIB	&	16.33	&	$	9791	\pm	1324	$	&	$	0.848	\pm	0.172	$	&	$	6438	\pm	871	$	&	$	1.876	\pm	0.381	$	&	$	12419	\pm	1680	$	&	$	0.667	\pm	0.135	$	\\
2010as	&	IIB	&	18.32	&	$	9083	\pm	1598	$	&	$	0.897	\pm	0.237	$	&	$	5671	\pm	997	$	&	$	2.402	\pm	0.635	$	&	$	12444	\pm	2189	$	&	$	0.644	\pm	0.170	$	\\
2010as	&	IIB	&	19.31	&	$	8218	\pm	1357	$	&	$	0.989	\pm	0.244	$	&	$	4991	\pm	824	$	&	$	3.172	\pm	0.782	$	&	$	11445	\pm	1889	$	&	$	0.688	\pm	0.169	$	\\
2010as	&	IIB	&	21.29	&	$	7224	\pm	1101	$	&	$	1.131	\pm	0.282	$	&	$	4492	\pm	685	$	&	$	3.922	\pm	0.977	$	&	$	9802	\pm	1494	$	&	$	0.782	\pm	0.195	$	\\
2010as	&	IIB	&	23.28	&	$	6445	\pm	959	$	&	$	1.272	\pm	0.332	$	&	$	4017	\pm	598	$	&	$	5.077	\pm	1.326	$	&	$	8741	\pm	1301	$	&	$	0.851	\pm	0.222	$	\\
2010as	&	IIB	&	25.26	&	$	6087	\pm	917	$	&	$	1.289	\pm	0.352	$	&	$	3763	\pm	567	$	&	$	5.772	\pm	1.574	$	&	$	8352	\pm	1259	$	&	$	0.836	\pm	0.228	$	\\
2010as	&	IIB	&	30.23	&	$	5375	\pm	623	$	&	$	1.363	\pm	0.333	$	&	$	3631	\pm	421	$	&	$	5.069	\pm	1.240	$	&	$	6698	\pm	776	$	&	$	0.964	\pm	0.236	$	\\
2010as	&	IIB	&	34.20	&	$	4906	\pm	572	$	&	$	1.468	\pm	0.382	$	&	$	3295	\pm	384	$	&	$	6.368	\pm	1.656	$	&	$	6157	\pm	718	$	&	$	0.999	\pm	0.260	$	\\
2010as	&	IIB	&	36.18	&	$	4824	\pm	483	$	&	$	1.427	\pm	0.338	$	&	$	3401	\pm	340	$	&	$	5.098	\pm	1.206	$	&	$	5794	\pm	580	$	&	$	1.035	\pm	0.245	$	\\
2011bm	&	Ic	&	28.96	&	$	5810	\pm	548	$	&	$	1.708	\pm	0.377	$	&	$	4685	\pm	442	$	&	$	3.273	\pm	0.723	$	&	$	6976	\pm	658	$	&	$	1.268	\pm	0.280	$	\\
2011bm	&	Ic	&	32.65	&	$	5502	\pm	494	$	&	$	1.772	\pm	0.425	$	&	$	4718	\pm	424	$	&	$	2.845	\pm	0.682	$	&	$	6440	\pm	579	$	&	$	1.323	\pm	0.317	$	\\
2011bm	&	Ic	&	35.60	&	$	5495	\pm	312	$	&	$	1.610	\pm	0.225	$	&	$	4718	\pm	267	$	&	$	2.576	\pm	0.360	$	&	$	6116	\pm	347	$	&	$	1.339	\pm	0.187	$	\\
2011bm	&	Ic	&	42.51	&	$	4854	\pm	343	$	&	$	1.810	\pm	0.347	$	&	$	4050	\pm	287	$	&	$	3.441	\pm	0.661	$	&	$	5563	\pm	394	$	&	$	1.394	\pm	0.268	$	\\
2011bm	&	Ic	&	48.40	&	$	4830	\pm	324	$	&	$	1.571	\pm	0.311	$	&	$	4198	\pm	282	$	&	$	2.559	\pm	0.507	$	&	$	5425	\pm	364	$	&	$	1.240	\pm	0.246	$	\\
2011bm	&	Ic	&	56.25	&	$	4844	\pm	253	$	&	$	1.252	\pm	0.176	$	&	$	4173	\pm	218	$	&	$	2.106	\pm	0.295	$	&	$	5358	\pm	280	$	&	$	1.032	\pm	0.145	$	\\
2011dh	&	IIB	&	6.21	&	$	7155	\pm	941	$	&	$	0.828	\pm	0.235	$	&	$	5338	\pm	702	$	&	$	1.718	\pm	0.488	$	&	$	9636	\pm	1268	$	&	$	0.558	\pm	0.159	$	\\
2011dh	&	IIB	&	14.30	&	$	6367	\pm	1064	$	&	$	1.054	\pm	0.404	$	&	$	4538	\pm	758	$	&	$	2.756	\pm	1.057	$	&	$	9054	\pm	1512	$	&	$	0.637	\pm	0.244	$	\\
2011dh	&	IIB	&	19.25	&	$	6320	\pm	859	$	&	$	0.955	\pm	0.304	$	&	$	4538	\pm	617	$	&	$	2.468	\pm	0.786	$	&	$	8702	\pm	1183	$	&	$	0.607	\pm	0.193	$	\\
2011dh	&	IIB	&	20.30	&	$	4686	\pm	499	$	&	$	2.126	\pm	0.742	$	&	$	4457	\pm	475	$	&	$	2.503	\pm	0.874	$	&	$	8833	\pm	940	$	&	$	0.570	\pm	0.199	$	\\
2011dh	&	IIB	&	27.24	&	$	4708	\pm	824	$	&	$	1.344	\pm	0.679	$	&	$	3291	\pm	576	$	&	$	5.239	\pm	2.646	$	&	$	6380	\pm	1116	$	&	$	0.758	\pm	0.383	$	\\
2011dh	&	IIB	&	41.14	&	$	3830	\pm	401	$	&	$	1.325	\pm	0.453	$	&	$	2721	\pm	285	$	&	$	6.615	\pm	2.264	$	&	$	4907	\pm	513	$	&	$	0.776	\pm	0.265	$	\\
2011dh	&	IIB	&	47.13	&	$	4016	\pm	543	$	&	$	1.031	\pm	2.114	$	&	$	2814	\pm	380	$	&	$	4.832	\pm	9.904	$	&	$	4863	\pm	657	$	&	$	0.680	\pm	1.393	$	\\
2011dh	&	IIB	&	48.15	&	$	4017	\pm	588	$	&	$	0.985	\pm	2.025	$	&	$	2752	\pm	403	$	&	$	5.109	\pm	10.503	$	&	$	4865	\pm	712	$	&	$	0.647	\pm	1.329	$	\\
2011dh	&	IIB	&	50.10	&	$	3880	\pm	452	$	&	$	0.988	\pm	0.417	$	&	$	2864	\pm	333	$	&	$	3.970	\pm	1.677	$	&	$	4989	\pm	581	$	&	$	0.565	\pm	0.239	$	\\
2011dh	&	IIB	&	57.09	&	$	3912	\pm	291	$	&	$	0.813	\pm	0.243	$	&	$	2996	\pm	223	$	&	$	2.722	\pm	0.813	$	&	$	4896	\pm	365	$	&	$	0.499	\pm	0.149	$	\\
2011dh	&	IIB	&	63.10	&	$	3953	\pm	358	$	&	$	0.725	\pm	0.247	$	&	$	3108	\pm	281	$	&	$	2.077	\pm	0.708	$	&	$	4833	\pm	438	$	&	$	0.461	\pm	0.157	$	\\
2011ei	&	IIB	&	15.77	&	$	12157	\pm	1593	$	&	$	0.199	\pm	0.039	$	&	$	9637	\pm	1263	$	&	$	0.277	\pm	0.054	$	&	$	14093	\pm	1847	$	&	$	0.171	\pm	0.033	$	\\
2011ei	&	IIB	&	16.68	&	$	11705	\pm	1109	$	&	$	0.202	\pm	0.033	$	&	$	9798	\pm	929	$	&	$	0.262	\pm	0.043	$	&	$	13445	\pm	1274	$	&	$	0.175	\pm	0.029	$	\\
2011ei	&	IIB	&	17.58	&	$	9270	\pm	1113	$	&	$	0.272	\pm	0.062	$	&	$	8505	\pm	1021	$	&	$	0.317	\pm	0.072	$	&	$	13081	\pm	1571	$	&	$	0.176	\pm	0.040	$	\\
2011ei	&	IIB	&	20.70	&	$	8504	\pm	1327	$	&	$	0.265	\pm	0.074	$	&	$	6234	\pm	973	$	&	$	0.506	\pm	0.140	$	&	$	11157	\pm	1741	$	&	$	0.192	\pm	0.053	$	\\
2011ei	&	IIB	&	26.59	&	$	6170	\pm	1114	$	&	$	0.354	\pm	0.140	$	&	$	4354	\pm	786	$	&	$	0.936	\pm	0.369	$	&	$	7746	\pm	1398	$	&	$	0.247	\pm	0.097	$	\\
2011ei	&	IIB	&	27.73	&	$	5922	\pm	767	$	&	$	0.352	\pm	0.101	$	&	$	4284	\pm	555	$	&	$	0.900	\pm	0.257	$	&	$	7091	\pm	918	$	&	$	0.266	\pm	0.076	$	\\
2011ei	&	IIB	&	30.72	&	$	5541	\pm	802	$	&	$	0.355	\pm	0.114	$	&	$	3731	\pm	540	$	&	$	1.206	\pm	0.387	$	&	$	6356	\pm	920	$	&	$	0.281	\pm	0.090	$	\\
2011ei	&	IIB	&	34.74	&	$	5493	\pm	294	$	&	$	0.324	\pm	0.049	$	&	$	3847	\pm	206	$	&	$	0.911	\pm	0.137	$	&	$	5598	\pm	300	$	&	$	0.313	\pm	0.047	$	\\
2011hs	&	IIB	&	10.47	&	$	6844	\pm	758	$	&	$	0.832	\pm	0.201	$	&	$	5005	\pm	554	$	&	$	1.824	\pm	0.442	$	&	$	8090	\pm	896	$	&	$	0.662	\pm	0.160	$	\\
2011hs	&	IIB	&	10.52	&	$	7055	\pm	412	$	&	$	0.795	\pm	0.145	$	&	$	5918	\pm	345	$	&	$	1.183	\pm	0.216	$	&	$	7619	\pm	444	$	&	$	0.715	\pm	0.130	$	\\
2011hs	&	IIB	&	11.46	&	$	6672	\pm	719	$	&	$	0.865	\pm	0.209	$	&	$	4912	\pm	529	$	&	$	1.896	\pm	0.458	$	&	$	7847	\pm	845	$	&	$	0.689	\pm	0.166	$	\\
2011hs	&	IIB	&	12.50	&	$	6276	\pm	829	$	&	$	0.999	\pm	0.284	$	&	$	4344	\pm	573	$	&	$	2.825	\pm	0.804	$	&	$	7769	\pm	1026	$	&	$	0.736	\pm	0.210	$	\\
2011hs	&	IIB	&	13.45	&	$	5890	\pm	825	$	&	$	1.155	\pm	0.358	$	&	$	3983	\pm	558	$	&	$	3.813	\pm	1.182	$	&	$	7475	\pm	1047	$	&	$	0.810	\pm	0.251	$	\\
2011hs	&	IIB	&	28.52	&	$	3581	\pm	546	$	&	$	0.821	\pm	0.358	$	&	$	2816	\pm	429	$	&	$	7.613	\pm	3.317	$	&	$	5004	\pm	763	$	&	$	0.996	\pm	0.434	$	\\
2011hs	&	IIB	&	31.40	&	$	3897	\pm	270	$	&	$	1.502	\pm	0.776	$	&	$	2972	\pm	206	$	&	$	5.100	\pm	2.635	$	&	$	4760	\pm	329	$	&	$	0.973	\pm	0.503	$	\\
2011hs	&	IIB	&	38.38	&	$	4048	\pm	352	$	&	$	1.076	\pm	0.305	$	&	$	3099	\pm	269	$	&	$	3.253	\pm	0.922	$	&	$	4629	\pm	402	$	&	$	0.804	\pm	0.228	$	\\
2011hs	&	IIB	&	42.32	&	$	4197	\pm	385	$	&	$	0.870	\pm	0.249	$	&	$	3173	\pm	291	$	&	$	2.674	\pm	0.767	$	&	$	4836	\pm	444	$	&	$	0.646	\pm	0.185	$	\\
2011hs	&	IIB	&	44.29	&	$	4239	\pm	377	$	&	$	0.811	\pm	0.226	$	&	$	3231	\pm	287	$	&	$	2.377	\pm	0.663	$	&	$	4863	\pm	432	$	&	$	0.609	\pm	0.170	$	\\
2012ap	&	IcBL	&	8.77	&	$	30930	\pm	33869	$	&	$	0.279	\pm	0.220	$	&	$	7771	\pm	8509	$	&	$	1.468	\pm	1.158	$	&	$	17835	\pm	4897	$	&	$	0.321	\pm	0.128	$	\\
2012ap	&	IcBL	&	9.73	&	$	28241	\pm	27934	$	&	$	0.282	\pm	0.208	$	&	$	7652	\pm	7569	$	&	$	1.422	\pm	1.050	$	&	$	13505	\pm	3048	$	&	$	0.409	\pm	0.150	$	\\
2012ap	&	IcBL	&	10.70	&	$	26601	\pm	26317	$	&	$	0.280	\pm	0.211	$	&	$	7245	\pm	7168	$	&	$	1.506	\pm	1.136	$	&	$	9871	\pm	1447	$	&	$	0.544	\pm	0.160	$	\\
2012ap	&	IcBL	&	11.75	&	$	23266	\pm	18859	$	&	$	0.294	\pm	0.193	$	&	$	7143	\pm	5790	$	&	$	1.453	\pm	0.953	$	&	$	8799	\pm	1097	$	&	$	0.588	\pm	0.163	$	\\
2012ap	&	IcBL	&	12.73	&	$	18355	\pm	12242	$	&	$	0.339	\pm	0.205	$	&	$	6317	\pm	4213	$	&	$	1.767	\pm	1.069	$	&	$	7054	\pm	1096	$	&	$	0.536	\pm	0.208	$	\\
2012ap	&	IcBL	&	13.70	&	$	15296	\pm	8591	$	&	$	0.384	\pm	0.217	$	&	$	5796	\pm	3255	$	&	$	2.043	\pm	1.157	$	&	$	6298	\pm	649	$	&	$	0.540	\pm	0.157	$	\\
2012ap	&	IcBL	&	14.61	&	$	12498	\pm	5299	$	&	$	0.459	\pm	0.226	$	&	$	5470	\pm	2319	$	&	$	2.230	\pm	1.097	$	&	$	-			$	&	$	-			$	\\
2012ap	&	IcBL	&	16.65	&	$	9772	\pm	2683	$	&	$	0.567	\pm	0.225	$	&	$	5226	\pm	1435	$	&	$	2.182	\pm	0.867	$	&	$	-			$	&	$	-			$	\\
2012ap	&	IcBL	&	17.65	&	$	8742	\pm	1973	$	&	$	0.649	\pm	0.239	$	&	$	5105	\pm	1152	$	&	$	2.208	\pm	0.812	$	&	$	-			$	&	$	-			$	\\
2012ap	&	IcBL	&	19.58	&	$	7783	\pm	1141	$	&	$	0.727	\pm	0.213	$	&	$	5289	\pm	775	$	&	$	1.795	\pm	0.527	$	&	$	-			$	&	$	-			$	\\
2012ap	&	IcBL	&	21.54	&	$	7257	\pm	905	$	&	$	0.758	\pm	0.210	$	&	$	5175	\pm	645	$	&	$	1.727	\pm	0.479	$	&	$	-			$	&	$	-			$	\\
2012ap	&	IcBL	&	32.52	&	$	5343	\pm	830	$	&	$	0.840	\pm	0.326	$	&	$	3509	\pm	545	$	&	$	3.619	\pm	1.403	$	&	$	-			$	&	$	-			$	\\
2012ap	&	IcBL	&	36.44	&	$	5386	\pm	555	$	&	$	0.703	\pm	0.205	$	&	$	4007	\pm	413	$	&	$	1.809	\pm	0.527	$	&	$	-			$	&	$	-			$	\\
1998bw	&	GRB	&	9.00	&	$	10581	\pm	2366	$	&	$	0.627		0.201	$	&	$	7181	\pm	1606	$	&	$	1.274	\pm	0.409	$	&	$	17938	\pm	4011	$	&	$	0.379	\pm	0.122	$	\\
1998bw	&	GRB	&	12.00	&	$	10635	\pm	2799	$	&	$	0.638		0.238	$	&	$	6801	\pm	1790	$	&	$	1.478	\pm	0.551	$	&	$	21353	\pm	5619	$	&	$	0.338	\pm	0.126	$	\\
1998bw	&	GRB	&	13.00	&	$	10550	\pm	2827	$	&	$	0.648		0.247	$	&	$	6695	\pm	1794	$	&	$	1.537	\pm	0.587	$	&	$	21731	\pm	5822	$	&	$	0.336	\pm	0.128	$	\\
1998bw	&	GRB	&	14.00	&	$	10346	\pm	2802	$	&	$	0.638		0.249	$	&	$	6526	\pm	1767	$	&	$	1.559	\pm	0.608	$	&	$	21731	\pm	5884	$	&	$	0.324	\pm	0.126	$	\\
1998bw	&	GRB	&	16.00	&	$	9853	\pm	2669	$	&	$	0.801		0.328	$	&	$	6185	\pm	1675	$	&	$	2.066	\pm	0.847	$	&	$	20968	\pm	5679	$	&	$	0.396	\pm	0.162	$	\\
1998bw	&	GRB	&	18.00	&	$	9167	\pm	2374	$	&	$	0.819		0.338	$	&	$	5820	\pm	1507	$	&	$	2.191	\pm	0.905	$	&	$	18633	\pm	4824	$	&	$	0.408	\pm	0.169	$	\\
1998bw	&	GRB	&	19.00	&	$	8766	\pm	2177	$	&	$	0.842		0.346	$	&	$	5635	\pm	1399	$	&	$	2.271	\pm	0.933	$	&	$	17044	\pm	4233	$	&	$	0.427	\pm	0.175	$	\\
1998bw	&	GRB	&	21.00	&	$	8059	\pm	1832	$	&	$	0.889		0.358	$	&	$	5318	\pm	1209	$	&	$	2.410	\pm	0.970	$	&	$	14400	\pm	3273	$	&	$	0.469	\pm	0.189	$	\\
1998bw	&	GRB	&	24.00	&	$	7113	\pm	1415	$	&	$	1.014		0.405	$	&	$	4873	\pm	970	$	&	$	2.768	\pm	1.106	$	&	$	11466	\pm	2282	$	&	$	0.561	\pm	0.224	$	\\
1998bw	&	GRB	&	26.00	&	$	6615	\pm	1201	$	&	$	1.087		0.429	$	&	$	4650	\pm	844	$	&	$	2.932	\pm	1.156	$	&	$	10055	\pm	1826	$	&	$	0.623	\pm	0.246	$	\\
1998bw	&	GRB	&	27.00	&	$	6397	\pm	1118	$	&	$	1.160		0.454	$	&	$	4542	\pm	794	$	&	$	3.123	\pm	1.224	$	&	$	9514	\pm	1663	$	&	$	0.672	\pm	0.263	$	\\
1998bw	&	GRB	&	28.00	&	$	6213	\pm	1037	$	&	$	1.251		0.488	$	&	$	4468	\pm	746	$	&	$	3.313	\pm	1.292	$	&	$	9010	\pm	1503	$	&	$	0.737	\pm	0.288	$	\\
2003dh	&	GRB	&	8.30	&	$	-			$	&	$	-			$	&	$	8688	\pm	434	$	&	$	0.883	\pm	0.133	$	&	$	-			$	&	$	-			$	\\
2003dh	&	GRB	&	9.10	&	$	-			$	&	$	-			$	&	$	7976	\pm	399	$	&	$	0.844	\pm	0.083	$	&	$	-			$	&	$	-			$	\\
2003dh	&	GRB	&	10.00	&	$	-			$	&	$	-			$	&	$	6875	\pm	344	$	&	$	1.307	\pm	0.086	$	&	$	-			$	&	$	-			$	\\
2003dh	&	GRB	&	22.10	&	$	-			$	&	$	-			$	&	$	5721	\pm	286	$	&	$	1.727	\pm	0.546	$	&	$	-			$	&	$	-			$	\\
2003lw	&	GRB	&	14.77	&	$	-			$	&	$	-			$	&	$	-			$	&	$	-			$	&	$	23996	\pm	1200	$	&	$	0.409	\pm	0.041	$	\\
2003lw	&	GRB	&	17.53	&	$	-			$	&	$	-			$	&	$	-			$	&	$	-			$	&	$	18602	\pm	930	$	&	$	0.448	\pm	0.045	$	\\
2003lw	&	GRB	&	18.63	&	$	-			$	&	$	-			$	&	$	-			$	&	$	-			$	&	$	28615	\pm	1431	$	&	$	0.295	\pm	0.029	$	\\
2006aj	&	GRB	&	7.61	&	$	-			$	&	$	-			$	&	$	7463	\pm	373	$	&	$	1.415	\pm	0.071	$	&	$	-			$	&	$	-			$	\\
2006aj	&	GRB	&	8.59	&	$	-			$	&	$	-			$	&	$	7613	\pm	381	$	&	$	1.295	\pm	0.065	$	&	$	-			$	&	$	-			$	\\
2006aj	&	GRB	&	9.54	&	$	-			$	&	$	-			$	&	$	6942	\pm	347	$	&	$	1.463	\pm	0.073	$	&	$	-			$	&	$	-			$	\\
2006aj	&	GRB	&	10.51	&	$	-			$	&	$	-			$	&	$	6389	\pm	319	$	&	$	1.633	\pm	0.082	$	&	$	-			$	&	$	-			$	\\
2006aj	&	GRB	&	11.48	&	$	-			$	&	$	-			$	&	$	5997	\pm	300	$	&	$	1.753	\pm	0.088	$	&	$	-			$	&	$	-			$	\\
2006aj	&	GRB	&	12.44	&	$	-			$	&	$	-			$	&	$	5461	\pm	273	$	&	$	2.054	\pm	0.103	$	&	$	-			$	&	$	-			$	\\
2006aj	&	GRB	&	13.41	&	$	-			$	&	$	-			$	&	$	5332	\pm	267	$	&	$	2.024	\pm	0.101	$	&	$	-			$	&	$	-			$	\\
2006aj	&	GRB	&	14.38	&	$	-			$	&	$	-			$	&	$	4655	\pm	233	$	&	$	2.819	\pm	0.141	$	&	$	-			$	&	$	-			$	\\
2006aj	&	GRB	&	15.35	&	$	-			$	&	$	-			$	&	$	4855	\pm	243	$	&	$	2.274	\pm	0.114	$	&	$	-			$	&	$	-			$	\\
2006aj	&	GRB	&	18.11	&	$	-			$	&	$	-			$	&	$	4213	\pm	211	$	&	$	2.695	\pm	0.135	$	&	$	-			$	&	$	-			$	\\
2006aj	&	GRB	&	19.21	&	$	-			$	&	$	-			$	&	$	3989	\pm	199	$	&	$	2.966	\pm	0.148	$	&	$	-			$	&	$	-			$	\\
2009nz	&	GRB	&	16.30	&	$	-			$	&	$	-			$	&	$	8959	\pm	448	$	&	$	0.906	\pm	0.080	$	&	$	-			$	&	$	-			$	\\
2010bh	&	GRB	&	13.68	&	$	-			$	&	$	-			$	&	$	-			$	&	$	-			$	&	$	11855	\pm	593	$	&	$	0.206	\pm	0.011	$	\\
2010bh	&	GRB	&	14.62	&	$	-			$	&	$	-			$	&	$	-			$	&	$	-			$	&	$	11559	\pm	578	$	&	$	0.200	\pm	0.012	$	\\
2010bh	&	GRB	&	18.44	&	$	-			$	&	$	-			$	&	$	-			$	&	$	-			$	&	$	8872	\pm	444	$	&	$	0.218	\pm	0.009	$	\\
2010bh	&	GRB	&	20.29	&	$	-			$	&	$	-			$	&	$	-			$	&	$	-			$	&	$	7964	\pm	398	$	&	$	0.212	\pm	0.008	$	\\
2010bh	&	GRB	&	24.08	&	$	-			$	&	$	-			$	&	$	-			$	&	$	-			$	&	$	6317	\pm	316	$	&	$	0.249	\pm	0.017	$	\\
2010bh	&	GRB	&	30.68	&	$	-			$	&	$	-			$	&	$	-			$	&	$	-			$	&	$	4313	\pm	216	$	&	$	0.315	\pm	0.031	$	\\
2012bz	&	GRB	&	6.79	&	$	-			$	&	$	-			$	&	$	6193	\pm	310	$	&	$	2.328	\pm	0.116	$	&	$	-			$	&	$	-			$	\\
2012bz	&	GRB	&	8.35	&	$	-			$	&	$	-			$	&	$	6449	\pm	322	$	&	$	2.044	\pm	0.102	$	&	$	-			$	&	$	-			$	\\
2012bz	&	GRB	&	10.64	&	$	-			$	&	$	-			$	&	$	5971	\pm	299	$	&	$	2.155	\pm	0.108	$	&	$	-			$	&	$	-			$	\\
2012bz	&	GRB	&	11.66	&	$	-			$	&	$	-			$	&	$	6210	\pm	311	$	&	$	1.987	\pm	0.099	$	&	$	-			$	&	$	-			$	\\
2012bz	&	GRB	&	13.71	&	$	-			$	&	$	-			$	&	$	6204	\pm	310	$	&	$	1.907	\pm	0.095	$	&	$	-			$	&	$	-			$	\\
2012bz	&	GRB	&	15.36	&	$	-			$	&	$	-			$	&	$	6186	\pm	309	$	&	$	1.824	\pm	0.091	$	&	$	-			$	&	$	-			$	\\
2012bz	&	GRB	&	17.66	&	$	-			$	&	$	-			$	&	$	5669	\pm	283	$	&	$	1.958	\pm	0.098	$	&	$	-			$	&	$	-			$	\\
2012bz	&	GRB	&	19.18	&	$	-			$	&	$	-			$	&	$	5906	\pm	295	$	&	$	1.601	\pm	0.080	$	&	$	-			$	&	$	-			$	\\
2012bz	&	GRB	&	20.79	&	$	-			$	&	$	-			$	&	$	5041	\pm	252	$	&	$	2.292	\pm	0.115	$	&	$	-			$	&	$	-			$	\\
2013dx	&	GRB	&	10.45	&	$	-			$	&	$	-			$	&	$	5924	\pm	296	$	&	$	1.814	\pm	0.038	$	&	$	-			$	&	$	-			$	\\
2013dx	&	GRB	&	11.30	&	$	-			$	&	$	-			$	&	$	5276	\pm	264	$	&	$	2.494	\pm	0.055	$	&	$	-			$	&	$	-			$	\\
2013dx	&	GRB	&	13.04	&	$	-			$	&	$	-			$	&	$	4920	\pm	246	$	&	$	2.975	\pm	0.073	$	&	$	-			$	&	$	-			$	\\
2013dx	&	GRB	&	13.08	&	$	-			$	&	$	-			$	&	$	5218	\pm	261	$	&	$	2.459	\pm	0.060	$	&	$	-			$	&	$	-			$	\\
2013dx	&	GRB	&	13.90	&	$	-			$	&	$	-			$	&	$	4822	\pm	241	$	&	$	3.230	\pm	0.083	$	&	$	-			$	&	$	-			$	\\
2013dx	&	GRB	&	16.50	&	$	-			$	&	$	-			$	&	$	3965	\pm	198	$	&	$	5.277	\pm	0.156	$	&	$	-			$	&	$	-			$	\\
2013dx	&	GRB	&	16.57	&	$	-			$	&	$	-			$	&	$	3897	\pm	195	$	&	$	5.330	\pm	0.158	$	&	$	-			$	&	$	-			$	\\
2013dx	&	GRB	&	18.32	&	$	-			$	&	$	-			$	&	$	3926	\pm	196	$	&	$	5.111	\pm	0.164	$	&	$	-			$	&	$	-			$	\\
2013dx	&	GRB	&	19.11	&	$	-			$	&	$	-			$	&	$	3726	\pm	186	$	&	$	5.648	\pm	0.187	$	&	$	-			$	&	$	-			$	\\
2013dx	&	GRB	&	22.69	&	$	-			$	&	$	-			$	&	$	3813	\pm	191	$	&	$	4.534	\pm	0.170	$	&	$	-			$	&	$	-			$	\\
2013dx	&	GRB	&	25.32	&	$	-			$	&	$	-			$	&	$	2993	\pm	150	$	&	$	10.714	\pm	0.434	$	&	$	-			$	&	$	-			$	\\
2013dx	&	GRB	&	28.83	&	$	-			$	&	$	-			$	&	$	3149	\pm	157	$	&	$	7.023	\pm	0.322	$	&	$	-			$	&	$	-			$	\\
2016jca	&	GRB	&	5.52	&	$	-			$	&	$	-			$	&	$	13977	\pm	1398	$	&	$	0.503	\pm	0.071	$	&	$	-			$	&	$	-			$	\\
2016jca	&	GRB	&	6.26	&	$	-			$	&	$	-			$	&	$	8949	\pm	895	$	&	$	0.936	\pm	0.132	$	&	$	-			$	&	$	-			$	\\
2016jca	&	GRB	&	7.03	&	$	-			$	&	$	-			$	&	$	8638	\pm	864	$	&	$	0.976	\pm	0.138	$	&	$	-			$	&	$	-			$	\\
2016jca	&	GRB	&	7.06	&	$	-			$	&	$	-			$	&	$	9542	\pm	954	$	&	$	0.804	\pm	0.114	$	&	$	-			$	&	$	-			$	\\
2016jca	&	GRB	&	7.83	&	$	-			$	&	$	-			$	&	$	7949	\pm	795	$	&	$	1.055	\pm	0.149	$	&	$	-			$	&	$	-			$	\\
2016jca	&	GRB	&	8.61	&	$	-			$	&	$	-			$	&	$	7060	\pm	706	$	&	$	1.207	\pm	0.171	$	&	$	-			$	&	$	-			$	\\
2016jca	&	GRB	&	9.33	&	$	-			$	&	$	-			$	&	$	6044	\pm	604	$	&	$	1.586	\pm	0.224	$	&	$	-			$	&	$	-			$	\\
2016jca	&	GRB	&	10.08	&	$	-			$	&	$	-			$	&	$	5645	\pm	565	$	&	$	1.706	\pm	0.241	$	&	$	-			$	&	$	-			$	\\
2016jca	&	GRB	&	10.13	&	$	-			$	&	$	-			$	&	$	6240	\pm	624	$	&	$	1.281	\pm	0.181	$	&	$	-			$	&	$	-			$	\\
2016jca	&	GRB	&	10.90	&	$	-			$	&	$	-			$	&	$	4905	\pm	490	$	&	$	2.279	\pm	0.322	$	&	$	-			$	&	$	-			$	\\
2016jca	&	GRB	&	11.61	&	$	-			$	&	$	-			$	&	$	4954	\pm	495	$	&	$	2.180	\pm	0.308	$	&	$	-			$	&	$	-			$	\\
2016jca	&	GRB	&	11.67	&	$	-			$	&	$	-			$	&	$	5320	\pm	532	$	&	$	1.755	\pm	0.248	$	&	$	-			$	&	$	-			$	\\
2016jca	&	GRB	&	13.96	&	$	-			$	&	$	-			$	&	$	3721	\pm	372	$	&	$	4.519	\pm	0.639	$	&	$	-			$	&	$	-			$	\\
2016jca	&	GRB	&	15.44	&	$	-			$	&	$	-			$	&	$	3501	\pm	350	$	&	$	5.049	\pm	0.714	$	&	$	-			$	&	$	-			$	\\
2016jca	&	GRB	&	16.16	&	$	-			$	&	$	-			$	&	$	3919	\pm	392	$	&	$	3.326	\pm	0.470	$	&	$	-			$	&	$	-			$	\\
2016jca	&	GRB	&	16.92	&	$	-			$	&	$	-			$	&	$	3425	\pm	343	$	&	$	4.796	\pm	0.678	$	&	$	-			$	&	$	-			$	\\
2016jca	&	GRB	&	19.27	&	$	-			$	&	$	-			$	&	$	3252	\pm	325	$	&	$	5.054	\pm	0.715	$	&	$	-			$	&	$	-			$	\\
2016jca	&	GRB	&	21.38	&	$	-			$	&	$	-			$	&	$	3641	\pm	364	$	&	$	2.991	\pm	0.423	$	&	$	-			$	&	$	-			$	\\
2016jca	&	GRB	&	22.93	&	$	-			$	&	$	-			$	&	$	3371	\pm	337	$	&	$	3.803	\pm	0.538	$	&	$	-			$	&	$	-			$	\\
2016jca	&	GRB	&	23.92	&	$	-			$	&	$	-			$	&	$	2305	\pm	231	$	&	$	20.521	\pm	2.902	$	&	$	-			$	&	$	-			$	\\
2016jca	&	GRB	&	25.93	&	$	-			$	&	$	-			$	&	$	3677	\pm	368	$	&	$	2.181	\pm	0.308	$	&	$	-			$	&	$	-			$	\\
\hline
\label{table:DF}
\end{longtable}

\bsp

\label{lastpage}

\end{document}